\documentclass[aip,jcp,reprint,amsmath,amssymb,letterpaper,natbib,floatfix,superscriptaddress]{revtex4-1}

\usepackage{bm}
\usepackage{epsfig}
\usepackage{amsmath}
\usepackage{graphicx}
\usepackage{amssymb}
\usepackage[normalem]{ulem}
\usepackage[usenames]{color}

\begin{document}

\newcommand{\kB}{k_\mathrm{B}}
\newcommand{\tw}{t_\mathrm{w}}
\newcommand{\ee}{\mathrm{e}}

\newcommand{\eb}{\epsilon_{\rm b}}
\newcommand{\ebT}{\eb/T}
\newcommand{\nmax}{{\cal N}_\mathrm{max}}
\newcommand{\state}{{\cal N}}

\newcommand{\nb}{n_\mathrm{b}}
\newcommand{\nopt}{n_\mathrm{opt}}

\def\beq{\begin{equation}}
\def\eeq{\end{equation}}
\def\bea{\begin{eqnarray}}
\def\eea{\end{eqnarray}}

\def\cal#1{\mathcal{#1}}
\def\eqq#1{Eq.~(\ref{#1})}
\def\eq#1{(\ref{#1})}
\def\av#1{\langle #1 \rangle}

\newcommand{\TT}{\mathcal{T}}  
\newcommand{\FF}{\mathcal{F}}  

\definecolor{Blue}{rgb}{0,0.0,1.0}
\definecolor{Red}{rgb}{1,0.0,0.0}
\newcommand{\comment}[1]{\textcolor{Red}{#1}}
\newcommand{\change}[1]{\textcolor{Blue}{#1}}

\title{Analyzing mechanisms and  microscopic reversibility of self-assembly}
\author{James Grant}
\affiliation{Department of Physics, University of Bath, Bath BA2 7AY, UK}
\author{Robert L. Jack\footnote{\texttt{R.Jack@bath.ac.uk}}}
\affiliation{Department of Physics, University of Bath, Bath BA2 7AY, UK}
\author{Stephen Whitelam\footnote{\texttt{swhitelam@lbl.gov}}}
\affiliation{Molecular Foundry, Lawrence Berkeley National Laboratory, 1 Cyclotron Road, Berkeley, CA 94720, USA}

\begin{abstract}
We use computer simulations to investigate self-assembly in a system of model chaperonin proteins, and in an Ising lattice gas. We discuss the mechanisms responsible for rapid and efficient assembly in these systems, and we use measurements of dynamical activity and assembly progress to compare their propensities for kinetic trapping. We use the analytic solution of a simple minimal model to illustrate the key features associated with such trapping, paying particular attention to the number of ways that particles can misbind. We discuss the relevance of our results for the design and control of self-assembly in general.
\end{abstract}

\maketitle

\section{Introduction}

In self-assembly~\cite{Whitesides2002-science,sol07}, particles combine spontaneously to form structures that can be closed, like capsids~\cite{Hagan2006} and DNA `origami'~\cite{Rothemund2006}, or extended, like filaments~\cite{Yang2010a}, sheets~\cite{Paavola2006,Li2007-chap} and
unusual crystals~\cite{Leunissen2005,Glotzer2009-cryst,chung2010self,Romano2010,Miller2010}. The possibility of exploiting assembly for technological ends has been discussed many times~\cite{Whitesides2002-science,sol07}, but to realize this possibility we need to develop the ability to predict and control the properties of experimental self-assembling systems in general. In particular, understanding how systems can be designed so as to assemble reliably and rapidly while avoiding kinetic traps remains a key challenge.

Effective dynamical assembly typically requires bond-making {\em and} bond-breaking events, so that assembling particles can avoid long-lived disordered structures and form the desired ordered one. The role of transient unbinding during self-assembly is understood at a qualitative level~\cite{Whitesides2002-reverse,
Hagan2006,Jack2007,Wilber2007,Rapaport2008,Whitelam2009-collective,Hagan2011-mallet,Klotsa2011}:
particles on the micro- and nanoscale can exploit thermal fluctuations in
order to sample a range of bound configurations as structures grow. Such fluctuations allow particles to break local bonds and escape the kinetic `traps' that result when misbound particles become frozen into place by the arrival of more material. The importance of such fluctuations is apparent from measurements, in computer simulations, of assembly yield as a function of particle binding strength. Typically, such curves are non-monotonic, with a decrease in yield at large binding strength due to the suppression of bond-breaking events~\cite{Wilber2007,Hagan2006,Jack2007,klix2010,whitelam2010control} (see Fig.~\ref{fig:yield}). However, while the roles of fluctuations and transient unbinding are clear at this qualitative level, it is not clear `how much' reversibility is required for effective self-assembly in a given system.

Here we address this question. We introduce a toy model of assembly whose analytic solution demonstrates a minimal set of requirements for kinetic trapping. We also consider computer simulations of two models of interacting particles. 
The first is an off-lattice, coarse-grained model~\cite{whitelam2007avoiding} of `chaperonin' proteins from which filament-like and sheet-like structures can assemble. 
The second is the two-dimensional lattice gas, whose separation into dense and dilute phases exhibits many of the characteristic features of self-assembly~\cite{Whitelam2009-collective,Hagan2011-mallet}. 
We discuss the assembly mechanisms in these models, and in particular identify whether assembly is more efficient when a single structure forms by nucleation and growth, or when multiple structures form simultaneously.  
We then consider the role of thermal fluctuations, comparing measurements of dynamical activity~\cite{Baiesi2009,Garrahan2007-prl} 
with the flux towards the assembled state.  For example, as chaperonin particles assemble into a close-packed sheet, 
they typically bind and unbind hundreds or thousands of times before attaining their final positions.  
We find that both the mechanism of assembly
and the dynamical activity indicate the effectiveness of a system in avoiding (or escaping from) kinetic traps,
and we discuss the relevance of these results for the design of self-assembling systems.

\section{Models and assembly yields}

\subsection{General considerations}

\begin{figure}
\includegraphics[width=8.4cm]{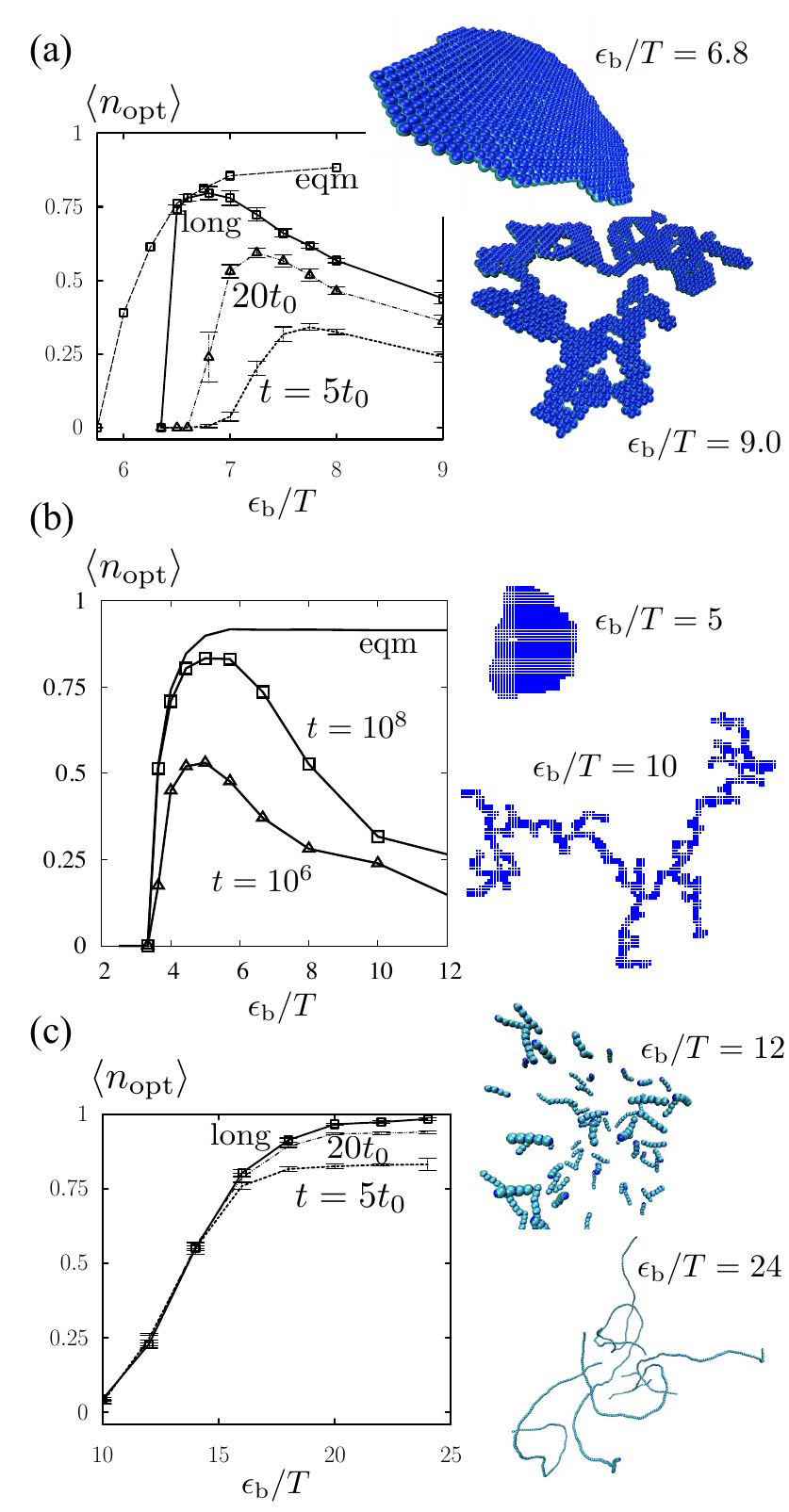}
\caption{Assembly yield $\langle \nopt \rangle$ versus binding strength $\epsilon_{\rm b}/T$, for various times and for
equilibrated systems.  We show representative snapshots of clusters at long times, for bond strengths indicated.
(a)~Sheet-forming chaperonin system ($\sigma=0.3$), (b)~lattice gas,  
(c)~filament-forming chaperonin system.  
In cases (a) and (b), dynamic yields at fixed time are non-monotonic in binding strength $\epsilon_{\rm b}/T$; 
in (c), yield is monotonic, reflecting the absence of kinetic trapping. 
Data marked `long' are taken from simulations lasting 300 hours of CPU time, rather than a fixed final time $t$.}
\label{fig:yield}
\end{figure}

A key aim of this article is to identify features that are conserved between different
self-assembling systems.  
To this end, we show results for three model systems, emphasising their common features
as well as some salient differences.
%
In all cases we initialise interacting particles in
disordered configurations and they evolve with diffusive dynamics towards low-energy 
thermally-equilibrated structures.  
For example, we will consider model chaperonin proteins 
that assemble into extended close-packed sheets
(full details are given in Sec.~\ref{sec:chap}).  
We define the `yield' of this assembly process to be 
the fraction of particles embedded in such close-packed sheets.  
To facilitate comparison between systems, we consistently use
$\eb/T$ to denote a dimensionless measure of the strength of interparticle bonds; we also
use  $\nopt$ to denote the assembly yield, defined as the
fraction of particles that are in `optimal' bonding environments.

Fig.~\ref{fig:yield}(a) shows results for the sheet-forming chaperonin system and Fig.~\ref{fig:yield}(b) shows
results for a two-dimensional lattice model where particles assemble into large close-packed clusters (see Sec.~\ref{sec:ising} for full details).
%
%
For these two systems, on these time scales,
$\nopt$ is large only in a narrow range of bond strength.  When bonds are too weak, the assembled structure
is not stable; when bonds are too strong, the system is vulnerable to kinetic trapping.   
We contrast this behavior with that of a different model of chaperonin proteins which assemble
into long filaments.
Fig.~\ref{fig:yield}(c) shows that this
process does not suffer kinetic trapping even when bonds are very strong: the
yield is monotonic in $\eb/T$.

\begin{figure*}
\includegraphics[width=16cm]{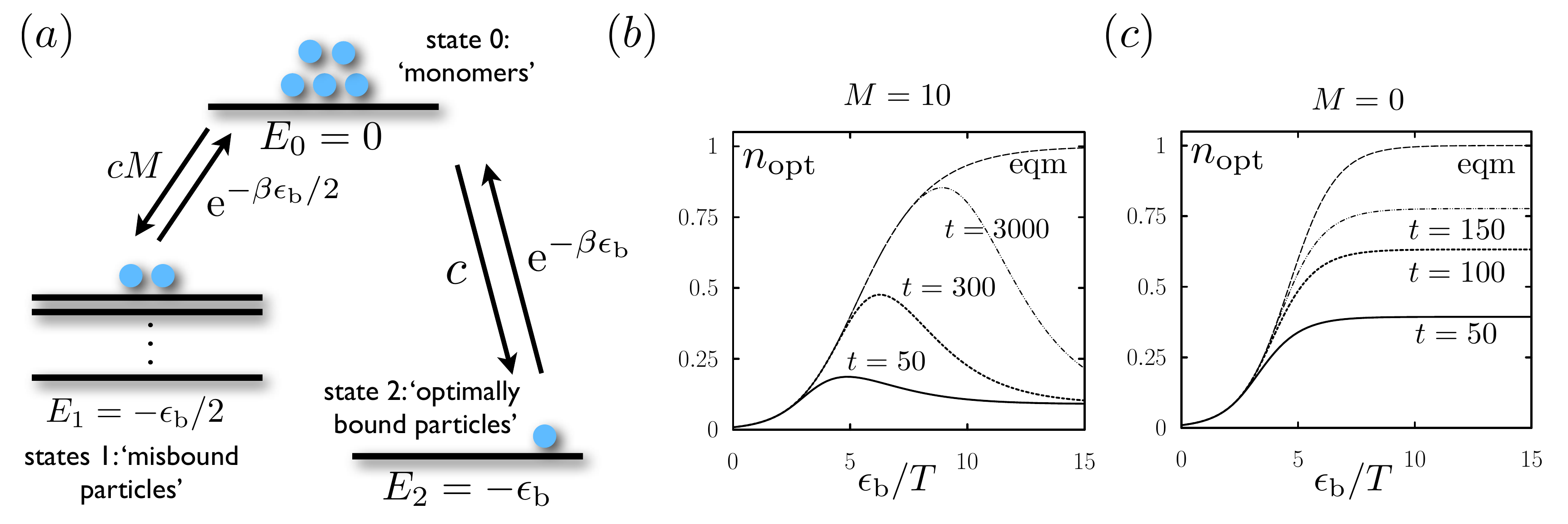}
\caption{\label{fig:trap} Analytic toy model of assembly demonstrating the requirements for kinetic trapping. (a) Particles transfer between the `monomer', `misbound' and `optimally bound' levels with the rates shown; $\eb$ is the particle binding strength; $c$ is a concentration variable (set to $10^{-2}$ in the other panels); and $M$ is the number of ways of misbinding. (b) When there exists the possibility of misbinding $(M>0)$, the dynamic yield is {\em non-monotonic} with $\eb$, because as $\eb$ increases 1) equilibrium yield {\em increases} but 2) the escape rate from misbound states {\em decreases}. (c) When misbinding is not possible ($M=0$), dynamic yield increases with binding strength. Similar behaviour is seen in computer models in Fig.~\ref{fig:yield}.}
\end{figure*}

\subsection{Chaperonin model}
\label{sec:chap}

Chaperonin proteins~\cite{Paavola2006,Li2007-chap} assemble \emph{in vitro} into a range of structures that include extended two-dimensional sheets and quasi one-dimensional filaments. Following Refs.~\cite{whitelam2007avoiding,Whitelam2009-chap}, we model chaperonins
 as hard spheres of diameter $2a$ equipped with orientation-dependent pairwise interactions that encourage either equator-to-equator or pole-to-pole binding (see Appendix~\ref{app:chap-model}). The anisotropic interactions have range $a/4$ and are characterized by a dimensionless bond strength $\ebT$. They also depend on a parameter $\sigma$ that determines how precisely two chaperonins must align before they receive an energetic reward: the smaller is $\sigma$, the more specific is the angular interaction. We simulated $N=1000$ chaperonins in periodically-replicated cubic boxes of side $L$. Chaperonins were present at a concentration of 0.82\% by volume (i.e. $\frac{4}{3} N \pi (a/L)^3 =0.0082$). 

In sheet-forming systems, particle interactions promote equator-to-equator binding. We focus
on a system with angular specificity parameter $\sigma=0.3$ (a fairly strict alignment criterion). We also contrast the behaviour of this system with that of a system possessing angular specificity parameter $\sigma=0.7$ (a more generous alignment criterion). At large bond strengths, equilibrium configurations of these systems contain a large close-packed planar sheet; for weak bonds the equilibrium is a dilute gas of free particles or small clusters. At the low concentrations studied, these systems do not form liquid phases or three-dimensional crystals. 

We also considered a filament-forming system whose interactions favour pole-to-pole binding. Its equilibrium state for large binding strength is a collection of long filaments.

For concreteness, we have selected particular values for parameters such as the specificity $\sigma$ and the volume fraction. Although there is a degree of arbitrariness in the particular values chosen, we find that the qualitative behaviour of the systems we consider here varies only weakly if we vary model parameters over a wide range of values. For instance, we do not find regimes in which the yields of chaperonin sheet formers or the lattice gas (Fig.~\ref{fig:yield}) vary monotonically with binding strength. Indeed, Fig.~\ref{fig:yield} shows that these systems exhibit similar qualitative trends, despite their differences in dimension, packing fraction, and the microscopic detail of their interactions. Similar behaviour has been observed in a range of other self-assembling systems~\cite{Whitesides2002-reverse,
Hagan2006,Jack2007,Wilber2007,Rapaport2008,Whitelam2009-collective,Hagan2011-mallet,Klotsa2011}. We are therefore
confident that our results are relevant for a range of self-assembling model systems; we would also expect similar
phenomenology to be reproduced in experiments.

We performed dynamic simulations, starting from well-mixed configurations, using the virtual-move Monte Carlo (MC) 
algorithm~\cite{whitelam2007avoiding} described in Ref.~\cite{whitelam2010approximating}. This algorithm approximates a diffusive dynamics by using potential energy gradients to generate both single-particle- and collective translations and rotations.  We define $\tau_\mathrm{B}$ as the mean time taken for an isolated particle to diffuse a length equal to its diameter (150 MC steps in our simulations). For later convenience we define $t_0 \equiv 10^5$ MC steps $\approx670 \tau_\mathrm{B}$. For the sheet-forming systems we sampled thermal equilibrium by starting from a large close-packed sheet inserted into a gas of monomers, and using local Monte Carlo moves supplemented by the nonlocal algorithm described in Ref.~\cite{whitelam2010control}.

To define the yield $\nopt$, we consider two particles $i$ and $j$ to be neighbours if their interaction energy $E_{ij} \leq -2T$.
For the sheet-forming model the optimal number of neighbours is $\nmax=6$; for the filament-forming model $\nmax=2$.
The yield $\nopt$ is the fraction of particles with this number of neighbours. We also define a normalised energy 
(`fraction of possible bonds')
\begin{equation}
\nb=-\frac{2E}{\nmax\eb},
\end{equation}
where $E$ is the total energy of the system.  Thus,
$\nb=0$ if no bonds are present, while $\nb=1$ if all particles are in optimal binding environments. 

The results shown in Fig~\ref{fig:yield} illustrate that the sheet-forming model suffers from kinetic trapping
when $\ebT$ is large, so that good assembly occurs only in an intermediate range of bond 
strengths~\footnote{We note also that dynamical trajectories of the chaperonin model equilibrate only within a narrow range of bond strengths. At small bond strengths, free energy barriers to sheet nucleation are large enough that they are not surmounted in direct simulations; at large bond strengths, disordered aggregates form and do not relax on timescales simulated.}.  On the other hand,
growing filaments in this model cannot become kinetically trapped: each particle can bind only at its north or south pole, and each of those two modes of binding permits the structure to be extended in an orderly manner. 
In this case, thermal fluctuations do not facilitate assembly, but instead break up long filaments and reduce yield. We note that assembly of filaments may still suffer from kinetic trapping if they have more internal structure than the simple strings
of particles considered here~\cite{Yang2010a,Fandrich2009}.

\subsection{Lattice gas}
\label{sec:ising}

We also consider the two-dimensional lattice gas, comprising $N$ particles on a square lattice of $V=L^2$ sites. Particles on nearest-neighbouring sites form bonds of energy $-\eb$; particles may not overlap. The system phase-separates
when bonds are strong, forming dense (liquid) and dilute (gas) phases. We work at density $\rho\equiv N/V=0.002$ for which the onset of phase separation (binodal)
is at $\ebT=3.2$~\cite{Baxter-book}.  We take $L=2048$ throughout.
Motivated by the characteristic non-monotonic yield shown in Fig.~\ref{fig:yield}(b), we draw
an analogy between this phase separation and the self-assembly observed in the chaperonin 
model~\cite{Whitelam2009-collective,Hagan2011-mallet}.  In the limit of large $\ebT$ we observe diffusion-limited cluster aggregation~\cite{Meakin1983}, an example of kinetic trapping that frustrates phase separation.

We again used an MC scheme with cluster moves in order to simulate the dynamics of Brownian particles dispersed in a solvent.
Our scheme is a variant of the `cleaving' algorithm of Ref.~\cite{whitelam2007avoiding}: In each MC move we select a seed particle, and begin to grow a cluster by adding to the seed, with probability $p_{\rm c}=1-e^{-\lambda\eb/T}$, each of its neighbouring particles. Here $\lambda=0.9$ is a parameter that controls the relative
likelihood of moving single particles as opposed to whole clusters.  This process of adding particles to the cluster is repeated recursively until no more particles are added. We then attempt to move the resulting cluster in a random direction.  We reject any moves that would lead to more than one particle on any site; otherwise we calculate the energy difference $\Delta E$ between the original and proposed configurations. The cluster is moved with probability $p_{\rm m}=p_{\rm a}/n^2$ where $n$ is the size of the cluster and $p_{\rm a}=\min(1,\ee^{-(1-\lambda)\Delta E/T})$ if $\Delta E\neq0$. When $\Delta E=0$ we take $p_{\rm a}=\alpha$ with $\alpha=0.9$. The factors $p_{\rm a}$, $p_{\rm m}$ and $p_c$ together ensure that the dynamics obey detailed balance and that clusters of $n$ particles diffuse with a rate proportional to $1/n$.  The parameters $\alpha$ and $\lambda$ are chosen for computational efficiency, and for consistency with our other studies of this model~\cite{grant-jack-prep}. An MC sweep comprises $N$ MC moves. The Brownian time for an isolated particle is $\tau_\mathrm{B} \approx 1.11 $~MC sweep.
Equilibrium conditions were probed using simulations that were initialised with a large assembled cluster.

Particles are considered to be neighbours if they are on 
adjacent lattice sites.  The optimal number of neighbours is $\nmax=4$, allowing 
$\nopt$ and $\nb$ to be defined as in Sec.~\ref{sec:chap}.  Fig.~\ref{fig:yield}(b) shows that the assembly yield
of this simple two-dimensional lattice model is qualitatively similar to that of the sheet-forming chaperonin model.
In what follows, we use comparisons between these systems to identify which assembly properties may be generalised
between models, and which are model-dependent.

\subsection{Schematic model of assembly and kinetic trapping}
\label{sec:Mstate}

To illustrate the physical origins of the behaviour in Fig.~\ref{fig:yield}, we introduce a toy model of self-assembly. We consider a large number of particles, each of which can inhabit any of three energy levels: a `monomer' level of energy 0, a `misbound' level of energy $-\eb/2$, and an `optimally bound' level of energy $-\eb$: see Fig.~\ref{fig:trap}(a). Particles begin in the monomer level, and  transfer into the bound levels
with the displayed rates. Here $c$ is a concentration-like variable, and $M$ is the degeneracy of the misbound level, which reflects the number of ways a particle can misbind. Particles escape from bound states with the Arrhenius-like rates shown. 

Denoting the unbound, misbound and optimally bound states by $0,1,2$ respectively, the 
model is described by a master equation
\beq
\frac{\mathrm{d}}{\mathrm{d}t}{{\bm P}}(t)= 
W {\bm P}(t),
\label{equ:master}
\eeq
where ${\bm P}(t) \equiv \left(P_0(t), P_1(t),P_2(t) \right)$; the variable $P_i(t)$ is the probability that a particle resides 
in state $i$ at time $t$; and the matrix $W$ is
\begin{equation} W =
\left(   \begin{matrix} 
      -c (M+1) &  \alpha & \alpha^2 \\
      c M & -\alpha & 0 \\     
       c & 0 & -\alpha^2 
   \end{matrix} \right).
\end{equation}
We have defined $\alpha \equiv {\rm e}^{-\epsilon_{\rm b}/2T}$ for compactness of notation and we take Boltzmann's constant
 $k_\mathrm{B}=1$ throughout
this paper. 
The yield in this model is $\nopt\equiv P_2$.

 All particles start in the monomer state, so that Eq.~(\ref{equ:master}) is to be solved with the 
initial condition $\bm{P}(0)=(1,0,0)$.
The solution is obtained by matrix diagonalisation; details are given in  Appendix~\ref{app:minimal}.
In the long-time limit, ${\bm P}(t)$ converges to the equilibrium distribution $\bm{s}=\frac{1}{Z}(\alpha^2,cM\alpha,c)$ where
$Z=c+cM\alpha+\alpha^2$ is the partition function.  Thus the equilibrium (long-time) yield is $n_\mathrm{eq}=c/(c+cM\alpha+\alpha^2)$.%


Dynamic yields are shown in Fig.~\ref{fig:trap}(b,c).
The long-time yield $n_\mathrm{eq}$ {\em increases} as particle binding strength $\ebT$ increases. However, the escape rate $\alpha$
from the misbound state {\em decreases} as $\ebT$ increases, so that misbound particles take a long time to unbind and transfer to the bound state. 
As long as $M>0$, these two conflicting effects result in a yield $\nopt$ that at finite times {\em decreases} for large binding strength (Fig.~\ref{fig:trap}(b)). 
We show in Appendix \ref{app:minimal} that if $\alpha$ is small then reaching the equilibrium yield takes a time of order $(M+1)/\alpha$.
However, if $M=0$, i.e. there is no possibility of binding in a non-productive manner, then yield increases monotonically with binding strength (Fig.~\ref{fig:trap}(c)).

When $M>0$ the toy model reproduces the qualitative dependence of yield on time and bond strength shown in Figs.~\ref{fig:yield}(a,b). On the other hand, the behaviour shown in Fig.~\ref{fig:yield}(c) is reproduced by the toy model when $M=0$.
We see immediately the three requirements for a dynamic yield that is non-monotonic in particle binding strength: 1) equilibrium yield {\em increases} with increasing binding strength; 2) there exists the possibility of misbinding; and 3) the escape rate from misbound states {\em decreases} with increasing binding strength.

\section{Assembly mechanisms}
\label{sec:mech}

\begin{figure*}
\includegraphics[width=17.5cm]{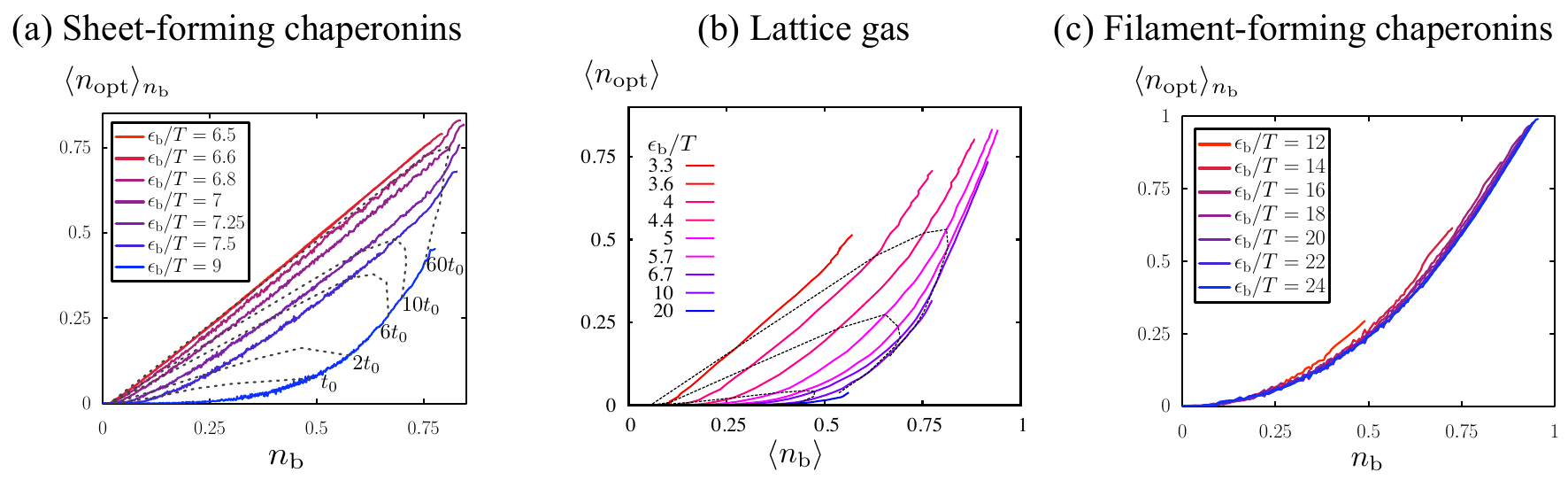}
\caption{Assembly quality $\nopt$ versus progress $\nb$ for (a) the chaperonin sheet-forming system ($\sigma=0.3$), 
(b) the lattice gas (isochrones are at $10^4$, $10^5$ and $10^6$ MC steps), and (c) the chaperonin filament-forming system.  We show data for a range of bond strengths $\epsilon_{\rm b}/T$,
as indicated. Time advances from bottom left to top right: dotted lines of constant time (isochrones) are drawn. In (a) the straight lines for the two highest temperatures indicate that assembly corresponds to the nucleation and growth of a single sheet; as temperature is lowered, multiple nucleation events are seen, and curves bend away from this line. 
Peak yield at long time is obtained (at $\epsilon_{\rm b}/T \approx 6.8$) slightly away from the single-sheet nucleation regime (peak yield is obtained even further from this regime for a sheet-forming system with a more generous angular binding criterion: see Fig.~\ref{fig:supp}(c)). 
Similar behaviour is seen in (b), although the nucleation regime is less pronounced. 
In (c), lowering temperature changes the assembly mechanism only slightly, and maximal yield at fixed time is always obtained for the lowest temperature considered.
}
\label{fig:mech}
\end{figure*}

We collate information about assembly mechanisms at different state points and in different models
 by plotting in Fig.~\ref{fig:mech} the normalized energy $\nb$ against the normalized 
number of optimally-bound particles $\nopt$ (which amounts to using energy as a measure of assembly progress). 
If the system contains large clusters of optimally bound particles then one expects $\nopt\approx\nb$.
However, particles on the cluster surfaces contribute to $\nb$ but not to $\nopt$ so one finds in general that $\nopt<\nb$.  For fixed $\nb$
the difference $\nb-\nopt$ is smallest when the system contains one large cluster, which has relatively few surface particles.  For
kinetically trapped states one typically finds $\nopt\ll\nb$, because few particles are in optimal environments.

In the chaperonin system there is a pronounced nucleation regime in which assembly proceeds by growth of a single
large cluster.  Since nucleation is a rare event, this regime is characterised by system-wide fluctuations.  However,
the assembly {\em mechanism} does not fluctuate, but is the same for all trajectories: a single sheet grows from a gas of particles 
(evidence for this assertion is given in Appendix~\ref{app:fluc}).  
In the parametric plots of Fig.~\ref{fig:mech},
this becomes clearest when we plot $\langle \nopt \rangle_{\nb}$, the assembly yield from multiple trajectories averaged over configurations 
with a given value of $\nb$. For the lattice gas system, the free energy barrier to nucleation is smaller, fluctuations between trajectories are less pronounced, and it is appropriate to take time as a parametric variable. We plot 
quantities averaged at constant time, $\langle \nopt(t) \rangle$ against $\langle \nb(t) \rangle$. We also show {\em isochrones}, lines connecting points of equal time (lattice gas), or points of equal average time (chaperonin systems).

Fig.~\ref{fig:mech} allows us to draw several conclusions about the assembly mechanism in these systems. In Fig.~\ref{fig:mech}(a) the nearly-straight lines at the two highest temperatures indicate that assembly corresponds to the nucleation and growth of a single sheet. As temperature is lowered, multiple nucleation events are seen, and curves bend away from this line.  (Since there are multiple growing sheets, the
fraction of bound particles located on cluster surfaces is larger, and $\nopt/\nb$ is lower, than in the single-sheet regime.) 
The maximal yield $\nopt$ at long times is obtained at $\epsilon_{\rm b}/T \approx 6.8$, slightly {\em away} from the single-sheet nucleation regime.
That is, while the ratio of surface to bulk particles is optimal in the single-sheet regime, the total number of assembled
particles increases with $\eb$ such that the yield continues to increase even as the surface-to-bulk ratio starts to fall.  This
competition between quality and quantity of assembled product was recently discussed in Ref.~\cite{Hagan2011-mallet}. Further from the single-sheet nucleation regime the surface-to-bulk effect dominates, and yield begins to decline. For very strong bonds, clusters become ramified, as illustrated in Fig.~\ref{fig:yield}, and yield is small.  
We note in passing that optimal assembly regime seems to take place near the spinodal line for phase separation,
since it is associated with a nucleation barrier that is just small enough for nucleation to cease to be a rare event -- the possibility
of controlling the nucleation barrier to achieve optimal assembly was discussed in~\cite{Miller2010}.

In Fig.~\ref{fig:mech}(b), we show data for the lattice gas model, which behaves similarly to the sheet-forming chaperonins:
maximal yield is obtained in a regime in which many clusters grow simultaneously, but too strong an interaction again impairs assembly. By contrast, the assembly mechanism in the filament-forming system is largely insensitive to bond strength: the main effect of increasing $\eb/T$ is that the system makes more progress along the reaction coordinate (Fig.~\ref{fig:mech}(c)). This again reflects the low propensity for kinetic trapping in this system.


\begin{figure*}
\includegraphics[width=\linewidth]{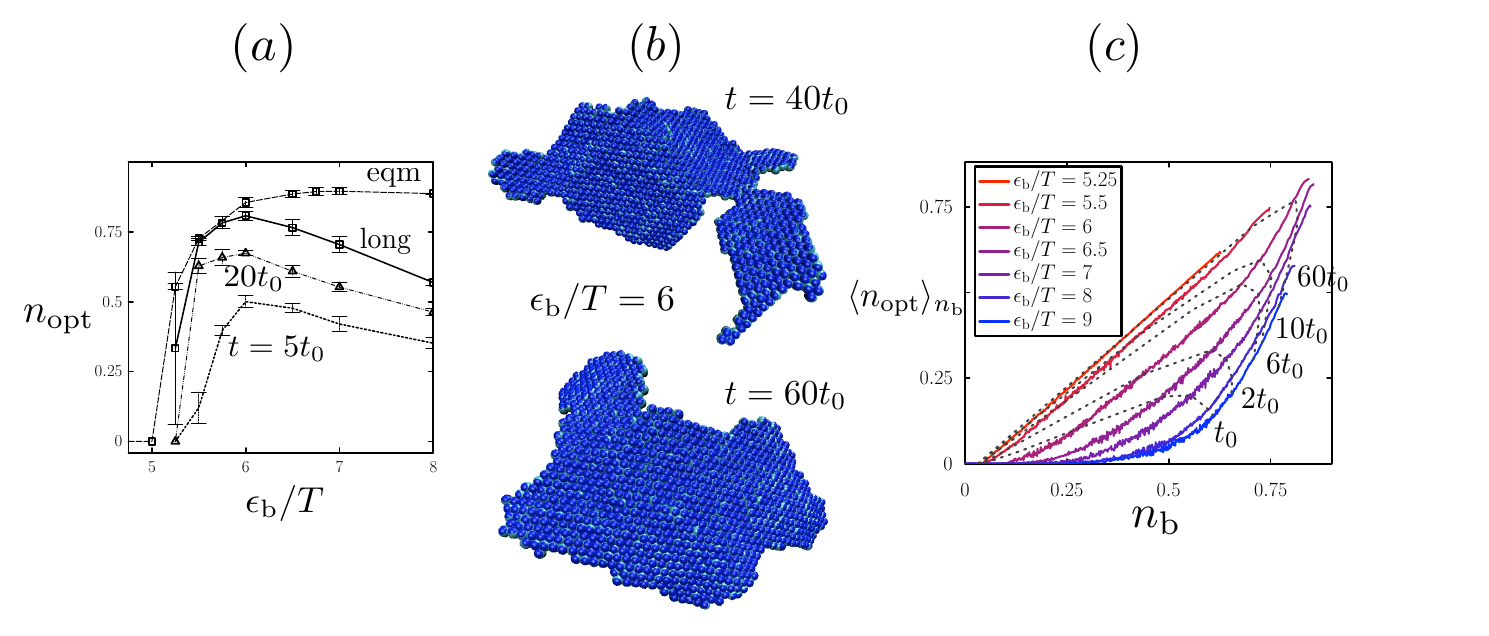}
\caption{Data for a chaperonin sheet-forming system with angular binding specificity $\sigma=0.7$ (less specific than the system
in Figs~\ref{fig:yield} and~\ref{fig:mech}).  (a) Yield; (b) fixed-time snapshots;
(c) assembly mechanism. While this system's behaviour is broadly similar to that shown in Figs.~\ref{fig:yield}(a) and \ref{fig:mech}(a), some details of its assembly are different. Notably, sheets can fuse and {\em heal} (b), eliminating joins formed by collisions and allowing particles to acquire optimal binding environments. As a result, peak yield (c) is found further from the regime of single-sheet nucleation than in Fig.~\ref{fig:mech}(a): this system can tolerate deeper supercooling than the one with $\sigma=0.3$.}
\label{fig:supp}
\end{figure*}

Finally, we note that despite their
different spatial dimensionality and binding geometry, the sheet-forming chaperonin model and the lattice gas
 show similar behaviour in the representations of
Figs.~\ref{fig:yield} and \ref{fig:mech}.  In both cases, assembly can take place through the nucleation and growth
of a single structure, but optimal yield occurs in the regime in which several clusters (sheets) grow simultaneously (see also~\cite{Klotsa2011}).
In Fig.~\ref{fig:supp} we show data for a chaperonin sheet-forming system with an angular binding specificity $(\sigma=0.7)$ 
more generous than that $(\sigma=0.3)$ studied in Figs.~\ref{fig:yield} and \ref{fig:mech}.
In particular, Fig,~\ref{fig:supp}(b) indicates that two intermediate-sized sheets may coalesce and heal into a single larger close-packed sheet.
This healing indicates that particles can escape kinetic traps. In Sec.~\ref{sec:outlook} we discuss this effect in the context
of assembly `forgivingness', the ability to recover an ordered product from a disordered intermediate state.

\section{Reversibility of binding}
\label{sec:rev}

\subsection{Everything put together (well) falls apart (transiently): Statistics of bond-breaking and bond-making}

\begin{figure}
\includegraphics[width=8.0cm]{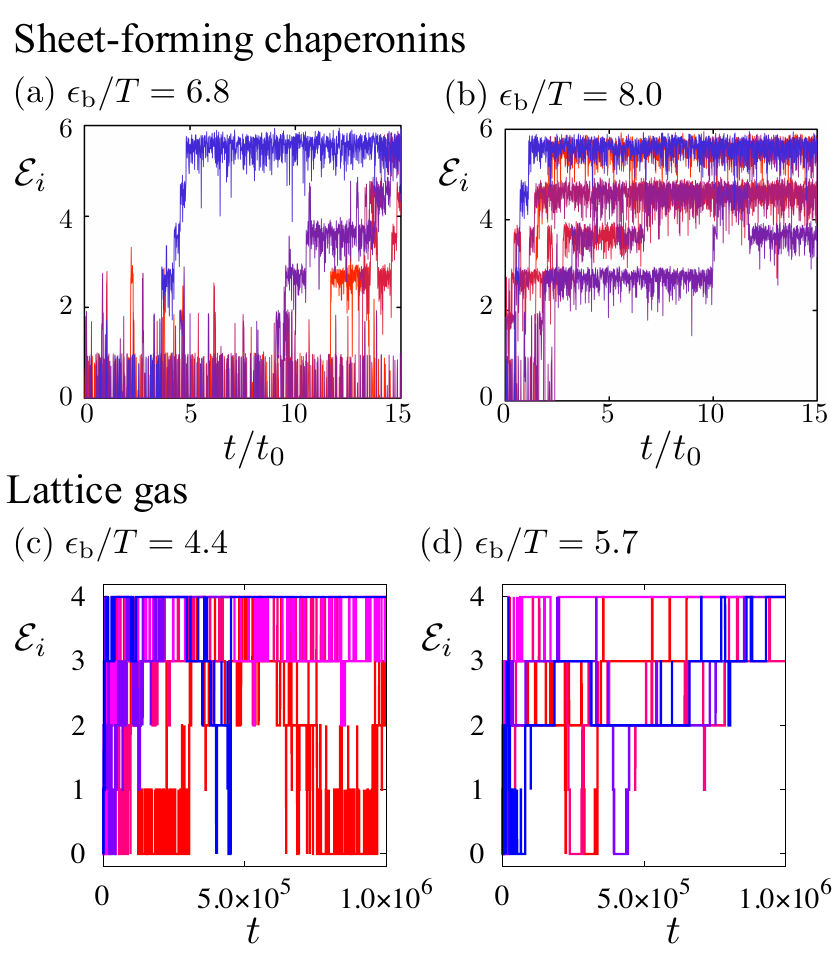}
\caption{
(a,b) Scaled energy ${\cal E}_i$ for each of 5 randomly-chosen sheet-forming chaperonin particles as a function of time $t$, 
for two different bond strengths and $\sigma=0.3$. (c,d) Similar data for lattice gas particles. 
Assembling particles bind {\em and} unbind, with unbinding being more frequent when bonds are weaker. The range of times shown is such that substantial assembly has occurred by the end of all trajectories.}
\label{fig:individual}
\end{figure}

As we have discussed (see e.g. Fig.~\ref{fig:trap}), non-monotonic yields such as those shown in Fig.~\ref{fig:yield} 
occur because assembling particles must break bonds that
are not compatible with the final ordered structure~\cite{Whitesides2002-reverse}.
In Fig.~\ref{fig:individual} we show the scaled energy ${\cal E}_i \equiv E_i /\eb$ of each of 5 randomly-chosen chaperonin sheet-formers as a function of the time $t$, for two different bond strengths.  
We show similar data for the lattice gas system. 
It is clear that assembling particles bind {\em and} unbind, and unbind more readily at the weaker bond strength. 
However, despite the clear link between bond-breaking events and good assembly, we possess little understanding of how many bond breakings are required in order to maintain effective assembly. 

To investigate this, we recorded for each particle the number of bound neighbours $\state_{\rm old}$ 
it possessed {\em before} each accepted MC move, and the number of bound neighbours $\state_{\rm new}$ 
it possessed {\em after} each accepted MC move.  If $\state_{\rm new} > \state_{\rm old}$ then we count a binding
event for this particle; if $\state_{\rm new} < \state_{\rm old}$ then we count an unbinding event~\footnote{We also tested a modified scheme in which particles making multiple bonds in one move contribute $\state_{\rm new} - \state_{\rm old}$ to $K_+$, etc. The results from this scheme and the one used in the main text are essentially indistinguishable.}. If $\state_{\rm new} = \state_{\rm old}$ then we assume that nothing happened to this particle (it might have gained and lost neighbours in equal number, but this happens so rarely in our simulations that we ignore it). We write $K_\pm$ to represent the total number of binding/unbinding events in a given time window of an assembly trajectory. We use these counts of binding and unbinding events to measure reversibility by separating them into 
time-reversal symmetric and asymmetric measures.  That is, averaging the numbers of events between times $0$ and $t$,
we define the {\em traffic} (or dynamical activity~\cite{Garrahan2007-prl,Baiesi2009}) as 
\begin{equation}
\cal{T}(t) \equiv \frac{ \langle K_+ \rangle + \langle K_- \rangle }N,
\end{equation}
and the {\em flux}  as
\begin{equation}
\cal{F}(t) \equiv \frac{\langle K_+ \rangle - \langle K_- \rangle}N.
\end{equation}
Traffic measures the total number of events per particle; 
flux measures the excess of binding over unbinding events per particle, and is a measure of the extent to which time-reversal symmetry is broken in the system.
For an equilibrated system (which is time-reversal symmetric), we have $\cal{F}(t)=0$ and
$\cal{T}(t)\propto t$.  For a system in which bonds never break, we have
$\cal{F}(t)=\cal{T}(t)$.  

We show typical results in Fig.~\ref{fig:flux-traffic}.  The maximal possible
flux in a system is approximately $\nmax$: flux increases in time in a similar way to 
$\langle \nb(t) \rangle$, because it quantifies the number of bonds in the system.  The flux therefore
saturates at long times, while the traffic continues to increase (events continue to happen in
the system even after it has equilibrated in the assembled state).

\subsection{Two steps forwards, one step back: quantifying reversibility}

\begin{figure}
\includegraphics[width=8.5cm]{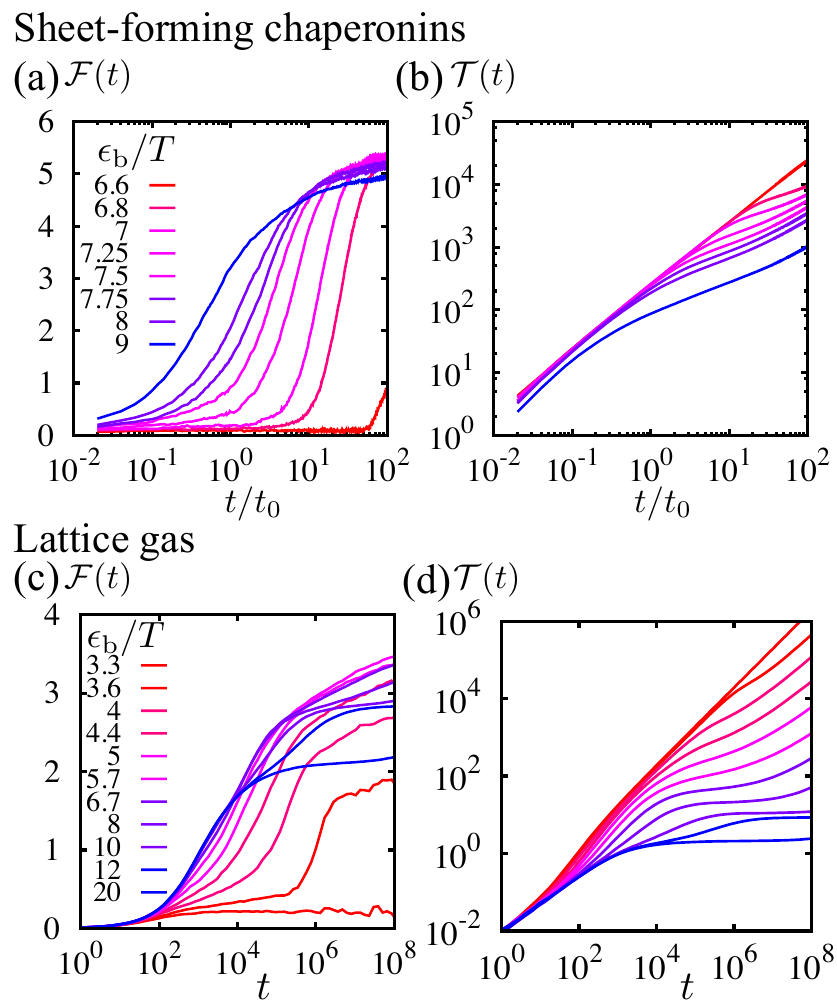}
\caption{(a,b) Flux and traffic measurements for the sheet-forming chaperonin system ($\sigma=0.3$);  (c,d)~Similar data for the lattice gas.
At fixed time, flux is non-monotonic in $\ebT$ (compare yield in Fig.~\ref{fig:yield}); but traffic
decreases with increasing $\ebT$, due to the role of the bond strength as an activation energy for bond-breaking.}
\label{fig:flux-traffic}
\end{figure}

Close inspection of Fig.~\ref{fig:flux-traffic} reveals that under optimal assembly conditions in the
sheet-forming system, particles eventually form on average about $5.5$ bonds, but participate in about $4000$ binding events (and so around $3994.5$ unbinding events). Lattice gas particles typically participate in approximately $2500$ binding and $2497$ unbinding events in order to achieve a net gain of $3$ bonds. In the filament-forming system, at the lowest temperature probed, 
particles participate in fewer than two events per bond formed. No reversibility is required in this case, because no misbinding can happen.

To interpret these results, it is useful to return to the toy model defined in Sec.~\ref{sec:Mstate}.  We assume
that reaching the `optimally bound' state results in 2 binding events, and reaching the `misbound' state results in
1 event (the idea is that optimally-bound particles typically have $\mathcal{N}_\mathrm{max}$ neighbours
while misbound ones have fewer than $\mathcal{N}_\mathrm{max}$; 
we take $\mathcal{N}_\mathrm{max}=2$).

\begin{figure}
\includegraphics[width=8.5cm]{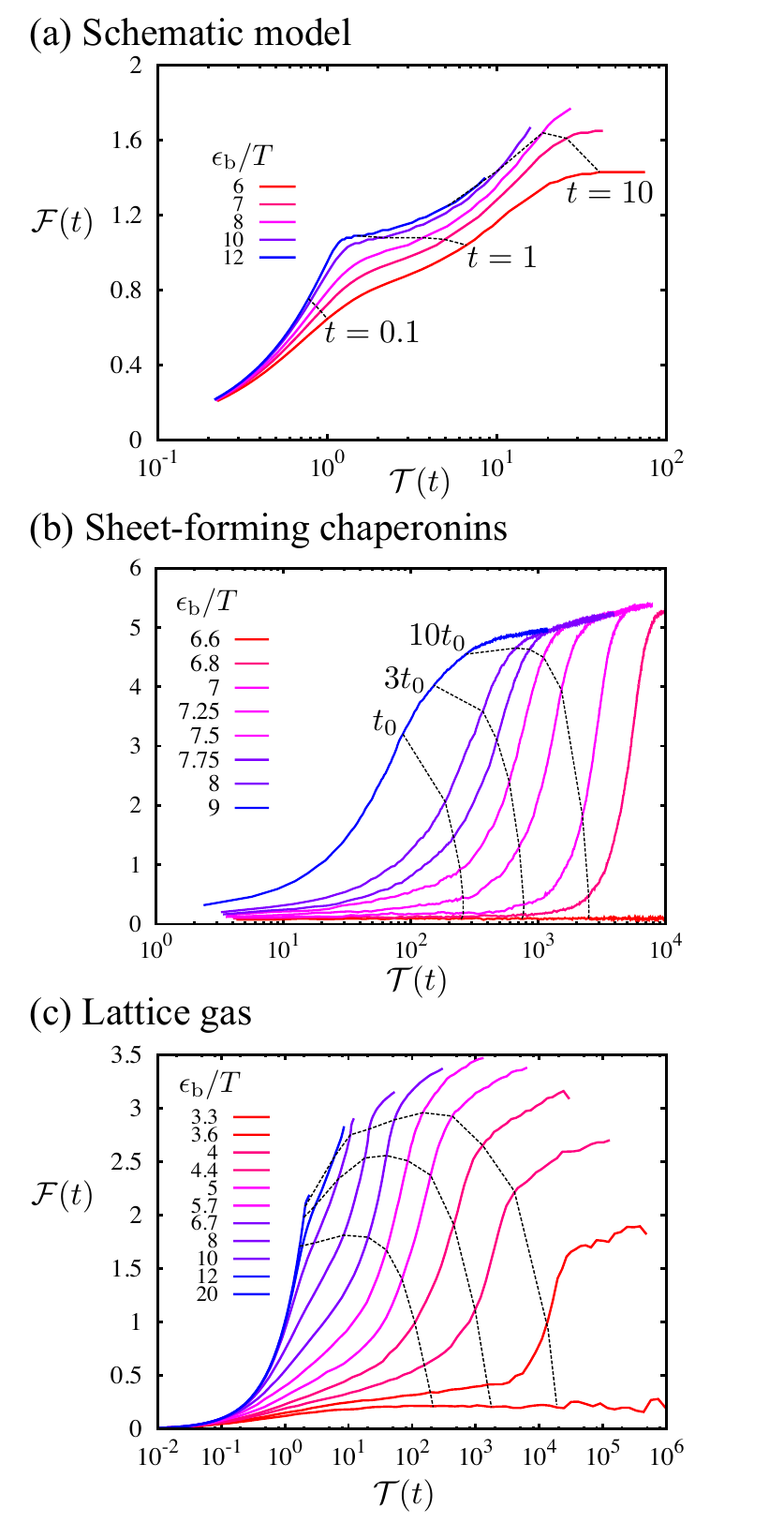}
\caption{
Parametric plots of flux and traffic during assembly, showing the role of traffic as a system
clock.  (a)~Schematic $M$-state model with $M=10$ and $c=0.01$. For large $\ebT$, 
assembly is complete after approximately $2(M+1)=22$ events.  (b)~Sheet-forming
chaperonin model ($\sigma=0.3$) with isochrones at the indicated times. 
(c)~Lattice gas model (isochrones
are at $10^4$, $10^5$ and $10^6$ MC steps).}
\label{fig:traffic-param}
\end{figure}

Assuming that $\ebT$ is large, it is useful to work
at leading order in $\alpha \equiv \ee^{-\eb/(2T)}$. The analysis is performed in Appendix~\ref{app:minimal}: here we summarise the main results.
In the limit of small $\alpha$ (and assuming $M>0$),
the toy model approaches the assembled state as $\nopt \sim 1-\ee^{-t/\tau}$ with $\tau\approx(M+1)/\alpha$.
There is a broad time window $\tau\ll t \ll \alpha^{-2}$ in which $\FF(t)\approx 2$ and $\TT(t)\approx 2(M+1)$.
Making a parametric plot of flux against traffic as in Fig.~\ref{fig:traffic-param}(a), one observes that for large $\eb$, 
the traffic plays the role of a clock, with the system reaching the assembled state when $\TT(t)\approx 2(M+1)$ (and $\TT(t)/\FF(t)=M+1$).  [If the limit of large $\eb$ has not yet been reached, 
the system reaches equilibrium 
at a value of $\TT(t)$ larger than $2(M+1)$.]

We show parametric plots of flux and traffic for the sheet-forming chaperonin model and the lattice gas in 
Figs.~\ref{fig:traffic-param}(b,c).
At a fixed value of traffic, flux is an increasing function of $\eb$, reflecting the role of $\eb$
as a driving force towards the assembled state.  But at fixed time, traffic is a decreasing function of $\eb$,
reflecting the role of $\eb$ in the activation energy for escaping from misbound states.  

\newcommand{\MM}{\tilde{M}}
\newcommand{\Meff}{M_\mathrm{eff}}

The parametric plots of Fig.~\ref{fig:traffic-param} allow comparison of reversibility of assembly between
different systems.  By comparison with the schematic model, we use this data to quantify systems' propensity for
kinetic trapping, as follows.  We calculate the ratio $\MM(t)=\TT(t)/\FF(t)$: under optimal assembly conditions
in the schematic model (large $\ebT$ and long time $t$) then $\MM(t)$ approaches $M+1$, the ratio of the number of
misbound and optimally bound states.  Given simulations of fixed length $t$ but varying $\ebT$, 
we define a parameter $\Meff(t)$ by evaluating $\MM(t)$ in the system with optimal $\ebT$.  In the schematic model,
$\Meff(t)\approx M+1$ as long as substantial assembly occurs before time $t$ for at least one value of $\ebT$.  

For our computer models, we obtain order-of-magnitude estimates of $\Meff$ as follows.  For the sheet-forming
chaperonins and times in the range $20-100t_0$, optimal assembly is in the range $7<\ebT<7.25$.  The flux is $\FF\approx5$
while the traffic is in the range $2000<\TT<7000$.  We infer that $\Meff$ lies in the range $400-1500$.  For the lattice
gas model and times in the range $10^7-10^8$ MC sweeps, optimal assembly is in the range $5<\ebT< 5.7$, the flux
is $\FF\approx3$ and the traffic in the range $500-5000$; the range for $\Meff$ is $200-2000$.  Given the large
overlap in estimates of $\Meff$ for lattice gas and chaperonin systems, we conclude that the propensity for kinetic trapping in these
two models are quite similar. For the sheet-forming chaperonins with $\sigma=0.7$ (see Fig.~\ref{fig:supp}) and taking 
time $20-100t_0$ we obtain
a range for $\Meff$ of $800-4000$, systematically larger than the value for the sheet-forming chaperonins with $\sigma=0.3$.
It may be that the less specific interaction potential offers more possibilities for disordered states, so that the system
requires more unbonding events in order to reach a final ordered structure.  For filament-forming chaperonins, optimal assembly
occurs at very large $\ebT$, for which $\TT\approx\FF$ and hence $\Meff=1$ (the analogous toy model has $M=\Meff-1=0$, as expected
since there
is no possibility for misbinding). 

We have emphasised the large uncertainties in the parameter $\Meff$: 
the model of Sec.~\ref{sec:Mstate} is a toy model of assembly, and one should not expect a direct mapping to more detailed computer models.
For example, the values we obtain for $M_\mathrm{eff}$ depend 
on the method used to identify neighbouring particles in the chaperonin model, and on the time at which flux and traffic
are measured.  Physically, the structures of the misbound states that cause kinetic trapping vary with time as assembly takes place, so describing these states with a single number $M_\mathrm{eff}$ is simplistic.  Nevertheless,
we argue that the parameter $M_\mathrm{eff}$ which we extract provides a useful estimate of the importance of kinetic trapping in these assembling systems.  Comparison of the values of $M_\mathrm{eff}$ emphasises the difference between sheet-forming and filament-forming chaperonins.  On the other hand, the difference between the sheet-forming chaperonins and the  assembling lattice gas model is very small, especially given the inherent uncertainties in estimating $M_\mathrm{eff}$.

In terms of effectiveness of assembly, we draw two main conclusions from the toy model.  Firstly, the time taken
to equilibrate depends strongly on the activation barrier for escape from misbound states, and is $\tau \sim (M+1) \ee^{-\eb/2T}$.
Thus, assembly is most rapid if the system possesses relatively weak bonds.  Secondly, the number of unbinding
events required to arrive at the assembled product depends on the number of misbound
states: this number reflects a system's propensity for trapping, and minimising $M$ provides a method for increasing
assembly quality.  Practical design rules for minimisation of $M$ remain an outstanding problem, but tuning the specificity
of inter-particle attractions~\cite{Whitelam2009-collective} might provide a route to minimizing this parameter.

\section{Outlook}
\label{sec:outlook}

Based on the analysis of this article, we draw two main conclusions.  In Section~\ref{sec:mech} we showed that the assembly mechanism assumed by classical nucleation theory (CNT), consisting of the growth of an isolated, compact cluster, typically operates when bonds are relatively weak. As bonds get stronger this simple picture no longer holds: multiple clusters
grow~\cite{Whitelam2009-collective,Klotsa2011}, and for very strong bonds cluster structures become ramified. We find that the competition between quality
and quantity of assembly~\cite{Hagan2011-mallet} results in optimal assembly happening away from the `CNT regime'. The extent to which this happens depends on the design of inter-component interactions (compare the sheet-forming systems with angular specificity $\sigma=0.7$ (Fig.~\ref{fig:supp}) 
with the data for $\sigma=0.3$ shown in the other Figures).

In Section~\ref{sec:rev} we demonstrated the importance to self-assembly of the reversibility of binding. In models
in which kinetic trapping is important, particles bind and unbind hundreds or thousands of times before finally
adopting their final positions in the assembled superstructure.  We associate the ratio of traffic and flux under conditions of optimal assembly
with a parameter $M_\mathrm{eff}$ that counts degeneracy of misbound states.  Large values of $M_\mathrm{eff}$
indicate that a system is prone to kinetic trapping; a system's bonds must be relatively weak in order to avoid such trapping.

These conclusions reinforce the importance of annealing if kinetic trapping is to be avoided.  
If departures from CNT at optimal assembly are large, then the system is effective
in annealing disordered clusters into well-formed products.  Similarly, if $M_\mathrm{eff}$ is small, the system
requires relatively few unbinding events in order to arrive at an assembled product.  Both
these measurements reflect the `forgivingness' of assembly, by which we mean the ability of a particles to escape from kinetic
traps and form an assembled product.  We believe that guidelines for improving forgivingness
are potentially useful in the design of self-assembly in general.  For the chaperonin sheet-formers that we considered, 
we found that the
version with reduced angular specificity seems to be the more forgiving of the two. Similarly, crystallisation 
tends to be most forgiving when interactions are relatively long-ranged; short-ranged interactions
more frequently lead to gelation or other forms of kinetic trapping~\cite{Sanz2008}.  Further simulation studies are needed in order to clarify the importance of microscopic parameters to the `forgivingness' of self-assembly, and to assess how typical numbers for `flux' and `traffic' compare to those seen in the model systems studied here. Ultimately, however, application of the ideas developed here requires the development of experimental systems in which the microscopic reversibility of self-assembling components can be quantified.

\begin{acknowledgments}
RLJ thanks Mike Hagan and Paddy Royall for many useful discussions.
This work was done as part of a User project at the Molecular Foundry, Lawrence Berkeley National Laboratory. JG and RLJ acknowledge financial support by the EPSRC through a doctoral training grant (to JG) and through grants EP/G038074/1 and EP/I003797/1 (to RLJ). SW was supported by the Director, Office of Science, Office of Basic Energy Sciences, of the U.S. Department of Energy under Contract No. DE-AC02--05CH11231.

\end{acknowledgments}

\begin{appendix}

\section{Minimal model of kinetic trapping}
\label{app:minimal}

In this appendix, we analyse the toy model introduced in Sec.~\ref{sec:Mstate}.
The master equation (\ref{equ:master}) can be solved exactly by matrix diagonalization: we write
$W=SDS^{-1}$ where $D$ is a diagonal matrix. The columns of $S$ are the right eigenvectors of $W$.  The solution 
is then $\bm{P}(t)=S\ee^{Dt}S^{-1}\bm{P}(0)$.  There is a zero eigenvalue of $W$ that corresponds to the steady state:
we denote the other two eigenvalues by $-\lambda_+$ and $-\lambda_-$ which are ordered as $0<\lambda_-<\lambda_+$.

\subsection{Assembly yield}
\label{app:M-yield}

The yield of assembly is $\nopt(t)\equiv P_2(t)$.  The
right eigenvector of $W$ corresponding to the equilibrium state is $\bm{s}=\frac{1}{Z}(\alpha^2,cM\alpha,c)$ where
$Z=c+cM\alpha+\alpha^2$ is the partition function.   Thus the equilibrium (long-time) yield is $n_\mathrm{eq}=\frac{c}{c+cM\alpha+\alpha^2}$
while for general $t$ the solution is of the form
\begin{equation}
\nopt(t) = n_\mathrm{eq}[1 - a\ee^{-\lambda_+ t} - b\ee^{-\lambda_- t}],
\label{equ:decay}
\end{equation}
where $a$ and $b$ are (positive) constants that depend on $\alpha$, $c$, and $M$, subject to $a+b=1$.

To gain physical intuition, it is convenient to assume that $\alpha$ is small.  In this case, we have
$\lambda_+ = c(M+1) + O(\alpha)$ while $\lambda_- = \frac{\alpha}{(M+1)}+O(\alpha^2)$.  Physically, the 
system forms bonds quickly (with rate $\lambda_+$), arriving in a state in which $P_2\approx \frac{1}{M+1}$ and
$P_1\approx \frac{M}{M+1}$.  There is then a slow relaxation (with rate $\lambda_-\ll 1$) in which
$P_2$ increases to the value $n_\mathrm{eq}\approx1$.  [Here and in the following, we use approximate
equalities to indicate that there are corrections at $O(\alpha)$.]  The slow relaxation to equilibrium involves
particles escaping from the misbound energy level, and therefore has an activated rate $\lambda_-\sim\ee^{-\eb/2T}$.
This gives rise to the non-monotonic yield plot shown in Fig.~\ref{fig:trap}(b).

When there is no possibility of misbinding (i.e. when $M=0$), the previous analysis holds but $b=0$
in (\ref{equ:decay}); the slow stage of
relaxation is irrelevant for the yield. In this case,
yield curves are monotonic with $\epsilon_{\rm b}/T$; see Fig.~\ref{fig:trap}(c).

\subsection{Flux and traffic}
\label{app:M-traffic}

To obtain time-averaged flux and traffic in this model, we notice that the average
number of transitions from state $1$ to state $0$ between times $0$ and $t$ is $K_{10} = \alpha \int_0^t\!\mathrm{d}t'\, P_1(t')$,
with similar results for transitions between other states.
For a full analysis of the statistics of the number of transitions between states in Markov processes,
see~\cite{Garrahan2009-kinks}.  If we assume that transitions between states $0$ and $1$ involve the making (or breaking)
of one bond while transitions between states $0$ and $2$ involve making or breaking of two bonds, we arrive at expressions for
the traffic and flux:
\begin{align}
\TT(t) &= \int_0^t\!\mathrm{d}t'\, [ c(M+2)P_0(t') + \alpha P_1(t') + 2\alpha^2 P_2(t') ],
\nonumber \\
\FF(t) &= \int_0^t\!\mathrm{d}t'\, [ -c(M+2)P_0(t') + \alpha P_1(t') + 2\alpha^2 P_2(t') ].
\end{align}
For the initial conditions used here, it may be readily shown from (\ref{equ:master}) that $\FF(t)=P_1(t)+2P_2(t)$, as required.

Using the solution $\bm{P}(t)$ given above and performing the time integral, 
one arrives at $$\TT(t) = (\bm{k}_+ + \bm{k}_-)^T S^{-1} D_\mathrm{int}(t) S \bm{P}(0),$$
where $\bm{k}_- = (0,\alpha,2\alpha^2)$, $\bm{k}_+ = (c(M+2),0,0)$ and $D_\mathrm{int}(t)$ is a diagonal matrix with elements
$(t,(1-\ee^{-\lambda_- t})/\lambda_-,(1-\ee^{\lambda_+ t})/\lambda_+)$.

We again analyse the limit of small $\alpha$.
For large times ($\lambda_-t\gg1$) the flux saturates at $2 + O(\alpha)$ while the traffic is
\begin{align}
\TT(t) \approx 2(M+1) + 2(M+2)\alpha^2t.
\end{align}
For large enough times, the second term dominates and the traffic increases linearly
with time, but if $\alpha^{-1} \ll t \ll \alpha^{-2}$ then traffic saturates at $2(M+1)$. 
This is the limit in which the number of unbinding events from the misbound state is large,
but unbinding events from the optimally-bound state are rare enough that they may be neglected.
The existence of such a limit is the basis for the extraction of the parameter $M_\mathrm{eff}$ discussed
in Sec.~\ref{sec:rev}.

%
%

\section{Inter-chaperonin potential}	
\label{app:chap-model}
Model chaperonins are hard spheres of diameter $2a$, equipped with an attractive pairwise interaction that operates only when the centres of two chaperonins lie within a distance $2a$ and  $2a+a/4$. Consider two chaperonins $i$ and $j$ that lie within this interaction range. Let ${\bm n}_i$ and ${\bm n}_j$ be unit vectors pointing from the centre of each chaperonin to its north pole, and let ${\bm r}_{ij}$ be the unit vector pointing from the centre of $i$ to the centre of $j$. Let $\phi_{ij}$ be the angle between the orientation vectors ${\bm n}_i$ and ${\bm n}_j$, and let $\theta_i$ be the angle between ${\bm n}_i$ and ${\bm r}_{ij}$ (and let $\theta_j$ be the angle between ${\bm n}_j$ and $-{\bm r}_{ij}$). Our `sticky equator' systems have orientational interaction
\beq
\epsilon_{\rm eq}=-\epsilon_{\rm b} \hat{C}_1(\phi_{ij};\sigma_{\rm align})C_0(\theta_i;\sigma_{\rm eq}) C_0(\theta_j;\sigma_{\rm eq}),
\label{eq:supp1}
\eeq
where
$C_{\alpha} \left(
\psi; \sigma\right) \equiv e^{- \left( \cos \psi-\alpha
\right)^2/\sigma^2 }$ rewards the alignment of angles $\psi$ and
$\cos^{-1}\alpha$. The parameter $\sigma$ determines the angular tolerance of this interaction. $\hat{
C}_{\alpha} \left( \psi; \sigma\right) \equiv C_{\alpha} \left( \psi;
\sigma \right) + C_{-\alpha} \left( \psi; \sigma \right)$ is this function's symmetrized counterpart. In \eqq{eq:supp1} the factors $C_0$ encourage orientation vectors to point perpendicular to the inter-chaperonin vector. The factor $ \hat{C}_1$ encourages orientation vectors to point parallel or antiparallel. For the sheet-forming system described in the main text we set $\sigma_{\rm align}=\sigma_{\rm eq}=0.3$. For the sheet-forming system described in Appendix B we set $\sigma_{\rm align}=\sigma_{\rm eq}=0.7$.

For the `sticky pole' system we choose the angular interaction
\beq
\epsilon_{\rm pol}=-\epsilon_{\rm b}\hat{C}_1(\phi_{ij};\sigma_{\rm align})\hat{C}_1(\theta_i;\sigma_{\rm pol})
\hat{C}_1(\theta_j;\sigma_{\rm pol}),
\eeq
whose three functions encourage alignment vectors to point parallel or antiparallel (function 1), and alignment vectors to point parallel or antiparallel to the inter-chaperonin vector (functions 2 and 3). We set $\sigma_{\rm align}=0.3$ and $\sigma_{\rm pol}=0.12$.

\begin{figure}
\includegraphics[width=6.5cm]{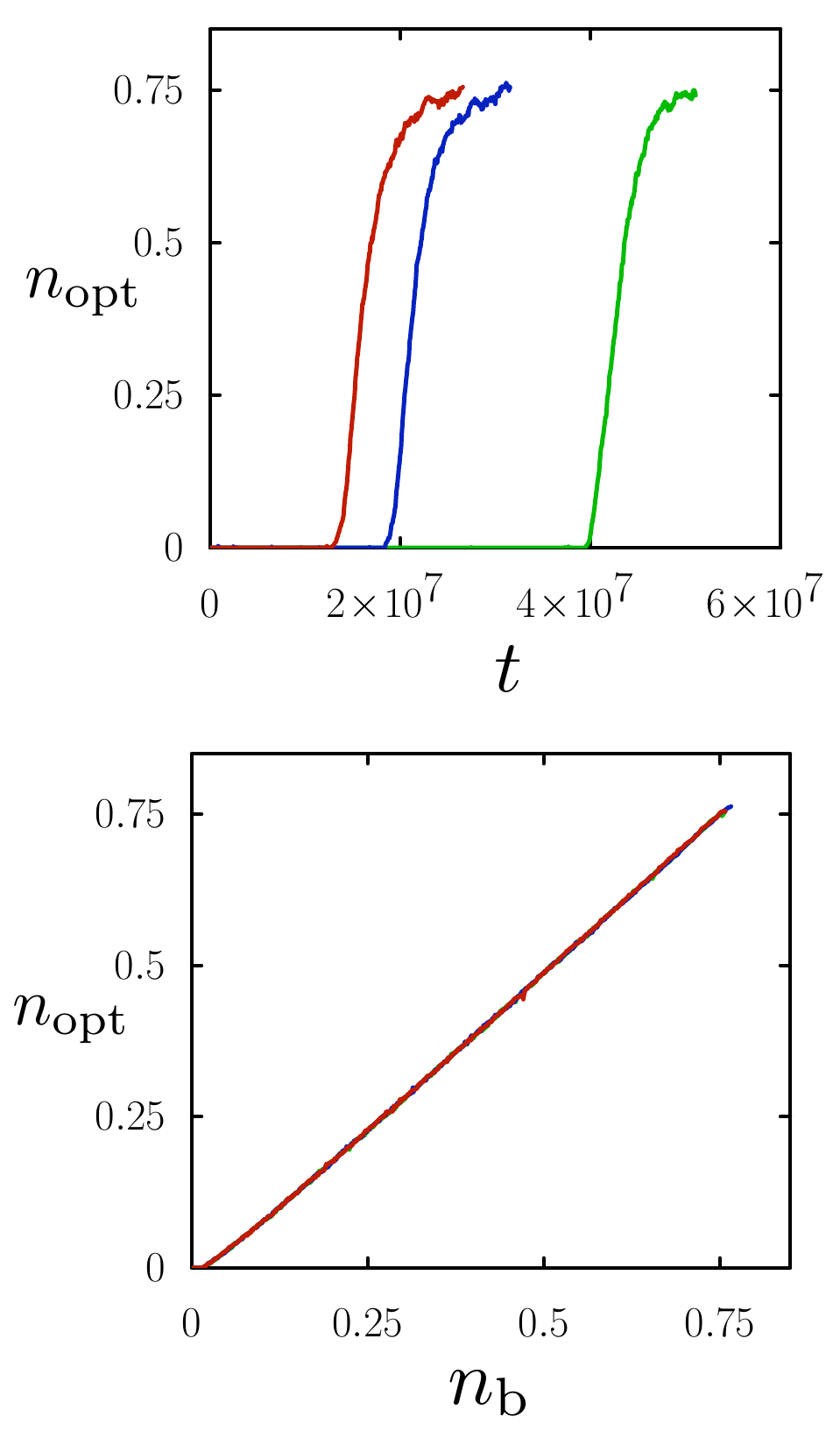}
\caption{System-wide fluctuations associated with nucleation can be controlled by using bond number (system energy) as a measure of assembly progress. (Top)~We show yield $n_{\rm opt}$ against time $t$ for three independent dynamical simulations of the sheet-forming model considered in the main text, for $\ebT=6.6$. Here assembly proceeds via nucleation and growth of a single sheet, and the characteristic time for appearance of the sheet is broadly distributed. (Bottom)~However, when bond number $n_{\rm b}$ is used as a measure of reaction progress, the data collapse. This collapse reveals that the assembly mechanism in all three trajectories is the same, and motivates the parametric plot of Fig.~\ref{fig:mech}.}
\label{fig:fluc}
\end{figure}

\section{Trajectory-to-trajectory fluctuations}
\label{app:fluc}

The self-assembly of sheets and lattice gas clusters reflects an underlying first-order phase transition, and can happen, roughly speaking, in one of two ways. Either a single critical nucleus appears in the system and grows by acquiring monomers, or many clusters of the new phase grow simultaneously and coalesce. Which of these mechanisms operates depends on the thermodynamic state and the system size (the latter is fixed in our simulations). The nucleation regime is characterised by large fluctuations: the randomly-distributed time at which the first critical nucleus appears strongly affects the behaviour of the whole system. In simulation studies, fluctuations associated with rare nucleation events lead to substantial differences in values of observables such as assembly yield $n_{\rm opt}(t)$ from run-to-run; the time-averaged yield $\langle \nopt(t) \rangle$ is usually not representative of the behaviour of any single trajectory. In order to deduce the assembly mechanism it is therefore useful to use the number of bonds in the system, rather than time, as a reaction coordinate. Fig.~\ref{fig:fluc} shows that data from different trajectories collapse in this representation: although the time to assembly varies significantly between trajectories, the assembly mechanism does not. 

\end{appendix}


\begin{thebibliography}{34}%
\makeatletter
\providecommand \@ifxundefined [1]{%
 \@ifx{#1\undefined}
}%
\providecommand \@ifnum [1]{%
 \ifnum #1\expandafter \@firstoftwo
 \else \expandafter \@secondoftwo
 \fi
}%
\providecommand \@ifx [1]{%
 \ifx #1\expandafter \@firstoftwo
 \else \expandafter \@secondoftwo
 \fi
}%
\providecommand \natexlab [1]{#1}%
\providecommand \enquote  [1]{``#1''}%
\providecommand \bibnamefont  [1]{#1}%
\providecommand \bibfnamefont [1]{#1}%
\providecommand \citenamefont [1]{#1}%
\providecommand \href@noop [0]{\@secondoftwo}%
\providecommand \href [0]{\begingroup \@sanitize@url \@href}%
\providecommand \@href[1]{\@@startlink{#1}\@@href}%
\providecommand \@@href[1]{\endgroup#1\@@endlink}%
\providecommand \@sanitize@url [0]{\catcode `\\12\catcode `\$12\catcode
  `\&12\catcode `\#12\catcode `\^12\catcode `\_12\catcode `\%12\relax}%
\providecommand \@@startlink[1]{}%
\providecommand \@@endlink[0]{}%
\providecommand \url  [0]{\begingroup\@sanitize@url \@url }%
\providecommand \@url [1]{\endgroup\@href {#1}{\urlprefix }}%
\providecommand \urlprefix  [0]{URL }%
\providecommand \Eprint [0]{\href }%
\@ifxundefined \urlstyle {%
  \providecommand \doi  [0]{\begingroup \@sanitize@url \@doi}%
  \providecommand \@doi [1]{\endgroup \@@startlink {\doibase
  #1}doi:\discretionary {}{}{}#1\@@endlink }%
}{%
  \providecommand \doi  [0]{doi:\discretionary{}{}{}\begingroup
  \urlstyle{rm}\Url }%
}%
\providecommand \doibase [0]{http://dx.doi.org/}%
\providecommand \Doi [0]{\begingroup \@sanitize@url \@Doi }%
\providecommand \@Doi  [1]{\endgroup\@@startlink{\doibase#1}\@@Doi}%
\providecommand \@@Doi [1]{#1\@@endlink}%
\providecommand \selectlanguage [0]{\@gobble}%
\providecommand \bibinfo  [0]{\@secondoftwo}%
\providecommand \bibfield  [0]{\@secondoftwo}%
\providecommand \translation [1]{[#1]}%
\providecommand \BibitemOpen [0]{}%
\providecommand \bibitemStop [0]{}%
\providecommand \bibitemNoStop [0]{.\EOS\space}%
\providecommand \EOS [0]{\spacefactor3000\relax}%
\providecommand \BibitemShut  [1]{\csname bibitem#1\endcsname}%
\bibitem [{\citenamefont {Whitesides}\ and\ \citenamefont
  {Grzybowski}(2002)}]{Whitesides2002-science}%
  \BibitemOpen
  \bibfield  {author} {\bibinfo {author} {\bibfnamefont {G.}~\bibnamefont
  {Whitesides}}\ and\ \bibinfo {author} {\bibfnamefont {B.}~\bibnamefont
  {Grzybowski}},\ }\href@noop {} {\bibfield  {journal} {\bibinfo  {journal}
  {Science},\ }\textbf {\bibinfo {volume} {295}},\ \bibinfo {pages} {2418}
  (\bibinfo {year} {2002})}\BibitemShut {NoStop}%
\bibitem [{\citenamefont {Glotzer}\ and\ \citenamefont
  {Solomon}(2007)}]{sol07}%
  \BibitemOpen
  \bibfield  {author} {\bibinfo {author} {\bibfnamefont {S.~C.}\ \bibnamefont
  {Glotzer}}\ and\ \bibinfo {author} {\bibfnamefont {M.~J.}\ \bibnamefont
  {Solomon}},\ }\href@noop {} {\bibfield  {journal} {\bibinfo  {journal}
  {Nature Mat.},\ }\textbf {\bibinfo {volume} {6}},\ \bibinfo {pages} {557}
  (\bibinfo {year} {2007})}\BibitemShut {NoStop}%
\bibitem [{\citenamefont {Hagan}\ and\ \citenamefont
  {Chandler}(2006)}]{Hagan2006}%
  \BibitemOpen
  \bibfield  {author} {\bibinfo {author} {\bibfnamefont {M.~F.}\ \bibnamefont
  {Hagan}}\ and\ \bibinfo {author} {\bibfnamefont {D.}~\bibnamefont
  {Chandler}},\ }\href@noop {} {\bibfield  {journal} {\bibinfo  {journal}
  {Biophys. J.},\ }\textbf {\bibinfo {volume} {91}},\ \bibinfo {pages} {42}
  (\bibinfo {year} {2006})}\BibitemShut {NoStop}%
\bibitem [{\citenamefont {Rothemund}(2006)}]{Rothemund2006}%
  \BibitemOpen
  \bibfield  {author} {\bibinfo {author} {\bibfnamefont {P.}~\bibnamefont
  {Rothemund}},\ }\href@noop {} {\bibfield  {journal} {\bibinfo  {journal}
  {Nature},\ }\textbf {\bibinfo {volume} {440}},\ \bibinfo {pages} {297}
  (\bibinfo {year} {2006})}\BibitemShut {NoStop}%
\bibitem [{\citenamefont {Yang}\ \emph {et~al.}(2010)\citenamefont {Yang},
  \citenamefont {Meyer},\ and\ \citenamefont {Hagan}}]{Yang2010a}%
  \BibitemOpen
  \bibfield  {author} {\bibinfo {author} {\bibfnamefont {Y.}~\bibnamefont
  {Yang}}, \bibinfo {author} {\bibfnamefont {R.}~\bibnamefont {Meyer}}, \ and\
  \bibinfo {author} {\bibfnamefont {M.~F.}\ \bibnamefont {Hagan}},\ }\href@noop
  {} {\bibfield  {journal} {\bibinfo  {journal} {Phys. Rev. Lett.},\ }\textbf
  {\bibinfo {volume} {104}},\ \bibinfo {pages} {258102} (\bibinfo {year}
  {2010})}\BibitemShut {NoStop}%
\bibitem [{\citenamefont {Paavola}\ \emph {et~al.}(2006)\citenamefont
  {Paavola}, \citenamefont {Chan}, \citenamefont {Li}, \citenamefont
  {Mazzarella}, \citenamefont {McMillan},\ and\ \citenamefont
  {Trent}}]{Paavola2006}%
  \BibitemOpen
  \bibfield  {author} {\bibinfo {author} {\bibfnamefont {C.}~\bibnamefont
  {Paavola}}, \bibinfo {author} {\bibfnamefont {S.}~\bibnamefont {Chan}},
  \bibinfo {author} {\bibfnamefont {Y.}~\bibnamefont {Li}}, \bibinfo {author}
  {\bibfnamefont {K.}~\bibnamefont {Mazzarella}}, \bibinfo {author}
  {\bibfnamefont {R.}~\bibnamefont {McMillan}}, \ and\ \bibinfo {author}
  {\bibfnamefont {J.}~\bibnamefont {Trent}},\ }\href@noop {} {\bibfield
  {journal} {\bibinfo  {journal} {Nanotechnology},\ }\textbf {\bibinfo {volume}
  {17}},\ \bibinfo {pages} {1171} (\bibinfo {year} {2006})}\BibitemShut
  {NoStop}%
\bibitem [{\citenamefont {Li}\ \emph {et~al.}(2007)\citenamefont {Li},
  \citenamefont {Paavola}, \citenamefont {Kagawa}, \citenamefont {Chan},\ and\
  \citenamefont {Trent}}]{Li2007-chap}%
  \BibitemOpen
  \bibfield  {author} {\bibinfo {author} {\bibfnamefont {Y.}~\bibnamefont
  {Li}}, \bibinfo {author} {\bibfnamefont {C.~D.}\ \bibnamefont {Paavola}},
  \bibinfo {author} {\bibfnamefont {H.}~\bibnamefont {Kagawa}}, \bibinfo
  {author} {\bibfnamefont {S.~L.}\ \bibnamefont {Chan}}, \ and\ \bibinfo
  {author} {\bibfnamefont {J.~D.}\ \bibnamefont {Trent}},\ }\href@noop {}
  {\bibfield  {journal} {\bibinfo  {journal} {Nanotechnology},\ }\textbf
  {\bibinfo {volume} {18}},\ \bibinfo {pages} {455101} (\bibinfo {year}
  {2007})}\BibitemShut {NoStop}%
\bibitem [{\citenamefont {Leunissen}\ \emph {et~al.}(2005)\citenamefont
  {Leunissen}, \citenamefont {Christova}, \citenamefont {Hynninen},
  \citenamefont {Royall}, \citenamefont {Campbell}, \citenamefont {Imhof},
  \citenamefont {Dijkstra}, \citenamefont {van Roij},\ and\ \citenamefont {van
  Blaaderen}}]{Leunissen2005}%
  \BibitemOpen
  \bibfield  {author} {\bibinfo {author} {\bibfnamefont {M.}~\bibnamefont
  {Leunissen}}, \bibinfo {author} {\bibfnamefont {C.}~\bibnamefont
  {Christova}}, \bibinfo {author} {\bibfnamefont {A.}~\bibnamefont {Hynninen}},
  \bibinfo {author} {\bibfnamefont {C.}~\bibnamefont {Royall}}, \bibinfo
  {author} {\bibfnamefont {A.}~\bibnamefont {Campbell}}, \bibinfo {author}
  {\bibfnamefont {A.}~\bibnamefont {Imhof}}, \bibinfo {author} {\bibfnamefont
  {M.}~\bibnamefont {Dijkstra}}, \bibinfo {author} {\bibfnamefont
  {R.}~\bibnamefont {van Roij}}, \ and\ \bibinfo {author} {\bibfnamefont
  {A.}~\bibnamefont {van Blaaderen}},\ }\Doi {DOI 10.1038/nature03946}
  {\bibfield  {journal} {\bibinfo  {journal} {Nature},\ }\textbf {\bibinfo
  {volume} {437}},\ \bibinfo {pages} {235} (\bibinfo {year}
  {2005})}\BibitemShut {NoStop}%
\bibitem [{\citenamefont {Iacovella}\ and\ \citenamefont
  {Glotzer}(2009)}]{Glotzer2009-cryst}%
  \BibitemOpen
  \bibfield  {author} {\bibinfo {author} {\bibfnamefont {C.~R.}\ \bibnamefont
  {Iacovella}}\ and\ \bibinfo {author} {\bibfnamefont {S.~C.}\ \bibnamefont
  {Glotzer}},\ }\href@noop {} {\bibfield  {journal} {\bibinfo  {journal} {Nano
  Lett.},\ }\textbf {\bibinfo {volume} {9}},\ \bibinfo {pages} {1206} (\bibinfo
  {year} {2009})}\BibitemShut {NoStop}%
\bibitem [{\citenamefont {Chung}\ \emph {et~al.}(2010)\citenamefont {Chung},
  \citenamefont {Shin}, \citenamefont {Bertozzi},\ and\ \citenamefont
  {De~Yoreo}}]{chung2010self}%
  \BibitemOpen
  \bibfield  {author} {\bibinfo {author} {\bibfnamefont {S.}~\bibnamefont
  {Chung}}, \bibinfo {author} {\bibfnamefont {S.}~\bibnamefont {Shin}},
  \bibinfo {author} {\bibfnamefont {C.}~\bibnamefont {Bertozzi}}, \ and\
  \bibinfo {author} {\bibfnamefont {J.}~\bibnamefont {De~Yoreo}},\ }\href@noop
  {} {\bibfield  {journal} {\bibinfo  {journal} {Proc. Natl. Acad. Sci.
  (USA)},\ }\textbf {\bibinfo {volume} {107}},\ \bibinfo {pages} {16536}
  (\bibinfo {year} {2010})}\BibitemShut {NoStop}%
\bibitem [{\citenamefont {Romano}\ \emph {et~al.}(2010)\citenamefont {Romano},
  \citenamefont {Sanz},\ and\ \citenamefont {Sciortino}}]{Romano2010}%
  \BibitemOpen
  \bibfield  {author} {\bibinfo {author} {\bibfnamefont {F.}~\bibnamefont
  {Romano}}, \bibinfo {author} {\bibfnamefont {E.}~\bibnamefont {Sanz}}, \ and\
  \bibinfo {author} {\bibfnamefont {F.}~\bibnamefont {Sciortino}},\ }\href@noop
  {} {\bibfield  {journal} {\bibinfo  {journal} {J. Chem. Phys.},\ }\textbf
  {\bibinfo {volume} {132}},\ \bibinfo {pages} {184501} (\bibinfo {year}
  {2010})},\ ISSN \bibinfo {issn} {00219606}\BibitemShut {NoStop}%
\bibitem [{\citenamefont {Miller}\ and\ \citenamefont
  {Cacciuto}(2010)}]{Miller2010}%
  \BibitemOpen
  \bibfield  {author} {\bibinfo {author} {\bibfnamefont {W.~L.}\ \bibnamefont
  {Miller}}\ and\ \bibinfo {author} {\bibfnamefont {A.}~\bibnamefont
  {Cacciuto}},\ }\href@noop {} {\bibfield  {journal} {\bibinfo  {journal} {J.
  Chem. Phys.},\ }\textbf {\bibinfo {volume} {133}},\ \bibinfo {pages} {234108}
  (\bibinfo {year} {2010})}\BibitemShut {NoStop}%
\bibitem [{\citenamefont {Whitesides}\ and\ \citenamefont
  {Boncheva}(2002)}]{Whitesides2002-reverse}%
  \BibitemOpen
  \bibfield  {author} {\bibinfo {author} {\bibfnamefont {G.~M.}\ \bibnamefont
  {Whitesides}}\ and\ \bibinfo {author} {\bibfnamefont {M.}~\bibnamefont
  {Boncheva}},\ }\href@noop {} {\bibfield  {journal} {\bibinfo  {journal}
  {Proc. Natl. Acad. Sci. (USA)},\ }\textbf {\bibinfo {volume} {99}},\ \bibinfo
  {pages} {4769} (\bibinfo {year} {2002})}\BibitemShut {NoStop}%
\bibitem [{\citenamefont {Jack}\ \emph {et~al.}(2007)\citenamefont {Jack},
  \citenamefont {Hagan},\ and\ \citenamefont {Chandler}}]{Jack2007}%
  \BibitemOpen
  \bibfield  {author} {\bibinfo {author} {\bibfnamefont {R.~L.}\ \bibnamefont
  {Jack}}, \bibinfo {author} {\bibfnamefont {M.~F.}\ \bibnamefont {Hagan}}, \
  and\ \bibinfo {author} {\bibfnamefont {D.}~\bibnamefont {Chandler}},\
  }\href@noop {} {\bibfield  {journal} {\bibinfo  {journal} {Phys. Rev. E},\
  }\textbf {\bibinfo {volume} {76}},\ \bibinfo {pages} {021119} (\bibinfo
  {year} {2007})}\BibitemShut {NoStop}%
\bibitem [{\citenamefont {Wilber}\ \emph {et~al.}(2007)\citenamefont {Wilber},
  \citenamefont {Doye}, \citenamefont {Louis}, \citenamefont {Noya},
  \citenamefont {Miller},\ and\ \citenamefont {Wong}}]{Wilber2007}%
  \BibitemOpen
  \bibfield  {author} {\bibinfo {author} {\bibfnamefont {A.~W.}\ \bibnamefont
  {Wilber}}, \bibinfo {author} {\bibfnamefont {J.~P.~K.}\ \bibnamefont {Doye}},
  \bibinfo {author} {\bibfnamefont {A.~A.}\ \bibnamefont {Louis}}, \bibinfo
  {author} {\bibfnamefont {E.~G.}\ \bibnamefont {Noya}}, \bibinfo {author}
  {\bibfnamefont {M.~A.}\ \bibnamefont {Miller}}, \ and\ \bibinfo {author}
  {\bibfnamefont {P.}~\bibnamefont {Wong}},\ }\href@noop {} {\bibfield
  {journal} {\bibinfo  {journal} {J. Chem. Phys.},\ }\textbf {\bibinfo {volume}
  {127}} (\bibinfo {year} {2007})}\BibitemShut {NoStop}%
\bibitem [{\citenamefont {Rapaport}(2008)}]{Rapaport2008}%
  \BibitemOpen
  \bibfield  {author} {\bibinfo {author} {\bibfnamefont {D.~C.}\ \bibnamefont
  {Rapaport}},\ }\href@noop {} {\bibfield  {journal} {\bibinfo  {journal}
  {Phys. Rev. Lett.},\ }\textbf {\bibinfo {volume} {101}},\ \bibinfo {pages}
  {186101} (\bibinfo {year} {2008})}\BibitemShut {NoStop}%
\bibitem [{\citenamefont {Whitelam}\ \emph
  {et~al.}(2009){\natexlab{a}}\citenamefont {Whitelam}, \citenamefont {Feng},
  \citenamefont {Hagan},\ and\ \citenamefont
  {Geissler}}]{Whitelam2009-collective}%
  \BibitemOpen
  \bibfield  {author} {\bibinfo {author} {\bibfnamefont {S.}~\bibnamefont
  {Whitelam}}, \bibinfo {author} {\bibfnamefont {E.~H.}\ \bibnamefont {Feng}},
  \bibinfo {author} {\bibfnamefont {M.~F.}\ \bibnamefont {Hagan}}, \ and\
  \bibinfo {author} {\bibfnamefont {P.~L.}\ \bibnamefont {Geissler}},\
  }\href@noop {} {\bibfield  {journal} {\bibinfo  {journal} {Soft Matter},\
  }\textbf {\bibinfo {volume} {5}},\ \bibinfo {pages} {1251} (\bibinfo {year}
  {2009}{\natexlab{a}})}\BibitemShut {NoStop}%
\bibitem [{\citenamefont {Hagan}\ \emph {et~al.}(2011)\citenamefont {Hagan},
  \citenamefont {Elrad},\ and\ \citenamefont {Jack}}]{Hagan2011-mallet}%
  \BibitemOpen
  \bibfield  {author} {\bibinfo {author} {\bibfnamefont {M.~F.}\ \bibnamefont
  {Hagan}}, \bibinfo {author} {\bibfnamefont {O.~M.}\ \bibnamefont {Elrad}}, \
  and\ \bibinfo {author} {\bibfnamefont {R.~L.}\ \bibnamefont {Jack}},\
  }\href@noop {} {\bibfield  {journal} {\bibinfo  {journal} {J. Chem. Phys.},\
  }\textbf {\bibinfo {volume} {135}},\ \bibinfo {pages} {104115} (\bibinfo
  {year} {2011})}\BibitemShut {NoStop}%
\bibitem [{\citenamefont {Klotsa}\ and\ \citenamefont
  {Jack}(2011)}]{Klotsa2011}%
  \BibitemOpen
  \bibfield  {author} {\bibinfo {author} {\bibfnamefont {D.}~\bibnamefont
  {Klotsa}}\ and\ \bibinfo {author} {\bibfnamefont {R.~L.}\ \bibnamefont
  {Jack}},\ }\href@noop {} {\bibfield  {journal} {\bibinfo  {journal} {Soft
  Matter},\ }\textbf {\bibinfo {volume} {6}},\ \bibinfo {pages} {6294}
  (\bibinfo {year} {2011})}\BibitemShut {NoStop}%
\bibitem [{\citenamefont {Klix}\ \emph {et~al.}(2010)\citenamefont {Klix},
  \citenamefont {Royall},\ and\ \citenamefont {Tanaka}}]{klix2010}%
  \BibitemOpen
  \bibfield  {author} {\bibinfo {author} {\bibfnamefont {C.~L.}\ \bibnamefont
  {Klix}}, \bibinfo {author} {\bibfnamefont {C.~P.}\ \bibnamefont {Royall}}, \
  and\ \bibinfo {author} {\bibfnamefont {H.}~\bibnamefont {Tanaka}},\ }\Doi
  {10.1103/PhysRevLett.104.165702} {\bibfield  {journal} {\bibinfo  {journal}
  {Phys. Rev. Lett.},\ }\textbf {\bibinfo {volume} {104}},\ \bibinfo {pages}
  {165702} (\bibinfo {year} {2010})}\BibitemShut {NoStop}%
\bibitem [{\citenamefont {Whitelam}(2010)}]{whitelam2010control}%
  \BibitemOpen
  \bibfield  {author} {\bibinfo {author} {\bibfnamefont {S.}~\bibnamefont
  {Whitelam}},\ }\Doi {10.1103/PhysRevLett.105.088102} {\bibfield  {journal}
  {\bibinfo  {journal} {Phys. Rev. Lett.},\ }\textbf {\bibinfo {volume}
  {105}},\ \bibinfo {pages} {088102} (\bibinfo {year} {2010})}\BibitemShut
  {NoStop}%
\bibitem [{\citenamefont {Whitelam}\ and\ \citenamefont
  {Geissler}(2007)}]{whitelam2007avoiding}%
  \BibitemOpen
  \bibfield  {author} {\bibinfo {author} {\bibfnamefont {S.}~\bibnamefont
  {Whitelam}}\ and\ \bibinfo {author} {\bibfnamefont {P.~L.}\ \bibnamefont
  {Geissler}},\ }\href@noop {} {\bibfield  {journal} {\bibinfo  {journal} {J.
  Chem. Phys.},\ }\textbf {\bibinfo {volume} {127}},\ \bibinfo {pages} {154101}
  (\bibinfo {year} {2007})}\BibitemShut {NoStop}%
\bibitem [{\citenamefont {Baiesi}\ \emph {et~al.}(2009)\citenamefont {Baiesi},
  \citenamefont {Maes},\ and\ \citenamefont {Wynants}}]{Baiesi2009}%
  \BibitemOpen
  \bibfield  {author} {\bibinfo {author} {\bibfnamefont {M.}~\bibnamefont
  {Baiesi}}, \bibinfo {author} {\bibfnamefont {C.}~\bibnamefont {Maes}}, \ and\
  \bibinfo {author} {\bibfnamefont {B.}~\bibnamefont {Wynants}},\ }\href@noop
  {} {\bibfield  {journal} {\bibinfo  {journal} {Phys. Rev. Lett.},\ }\textbf
  {\bibinfo {volume} {103}} (\bibinfo {year} {2009})}\BibitemShut {NoStop}%
\bibitem [{\citenamefont {Garrahan}\ \emph {et~al.}(2007)\citenamefont
  {Garrahan}, \citenamefont {Jack}, \citenamefont {Lecomte}, \citenamefont
  {Pitard}, \citenamefont {van Duijvendijk},\ and\ \citenamefont {van
  Wijland}}]{Garrahan2007-prl}%
  \BibitemOpen
  \bibfield  {author} {\bibinfo {author} {\bibfnamefont {J.~P.}\ \bibnamefont
  {Garrahan}}, \bibinfo {author} {\bibfnamefont {R.~L.}\ \bibnamefont {Jack}},
  \bibinfo {author} {\bibfnamefont {V.}~\bibnamefont {Lecomte}}, \bibinfo
  {author} {\bibfnamefont {E.}~\bibnamefont {Pitard}}, \bibinfo {author}
  {\bibfnamefont {K.}~\bibnamefont {van Duijvendijk}}, \ and\ \bibinfo {author}
  {\bibfnamefont {F.}~\bibnamefont {van Wijland}},\ }\Doi {ARTN} {\bibfield
  {journal} {\bibinfo  {journal} {Phys. Rev. Lett.},\ }\textbf {\bibinfo
  {volume} {98}},\ \bibinfo {pages} {195702} (\bibinfo {year}
  {2007})}\BibitemShut {NoStop}%
\bibitem [{\citenamefont {Whitelam}\ \emph
  {et~al.}(2009){\natexlab{b}}\citenamefont {Whitelam}, \citenamefont {Rogers},
  \citenamefont {Pasqua}, \citenamefont {Paavola}, \citenamefont {Trent},\ and\
  \citenamefont {Geissler}}]{Whitelam2009-chap}%
  \BibitemOpen
  \bibfield  {author} {\bibinfo {author} {\bibfnamefont {S.}~\bibnamefont
  {Whitelam}}, \bibinfo {author} {\bibfnamefont {C.}~\bibnamefont {Rogers}},
  \bibinfo {author} {\bibfnamefont {A.}~\bibnamefont {Pasqua}}, \bibinfo
  {author} {\bibfnamefont {C.}~\bibnamefont {Paavola}}, \bibinfo {author}
  {\bibfnamefont {J.}~\bibnamefont {Trent}}, \ and\ \bibinfo {author}
  {\bibfnamefont {P.~L.}\ \bibnamefont {Geissler}},\ }\href@noop {} {\bibfield
  {journal} {\bibinfo  {journal} {Nano Letters},\ }\textbf {\bibinfo {volume}
  {9}},\ \bibinfo {pages} {292} (\bibinfo {year}
  {2009}{\natexlab{b}})}\BibitemShut {NoStop}%
\bibitem [{\citenamefont {Whitelam}(2011)}]{whitelam2010approximating}%
  \BibitemOpen
  \bibfield  {author} {\bibinfo {author} {\bibfnamefont {S.}~\bibnamefont
  {Whitelam}},\ }\href@noop {} {\bibfield  {journal} {\bibinfo  {journal}
  {arXiv:1009.2008, Molecular Simulation, in press}} (\bibinfo {year}
  {2011})}\BibitemShut {NoStop}%
\bibitem [{Note1()}]{Note1}%
  \BibitemOpen
  \bibinfo {note} {We note also that dynamical trajectories of the chaperonin
  model equilibrate only within a narrow range of bond strengths. At small bond
  strengths, free energy barriers to sheet nucleation are large enough that
  they are not surmounted in direct simulations; at large bond strengths,
  disordered aggregates form and do not relax on timescales
  simulated.}\BibitemShut {Stop}%
\bibitem [{\citenamefont {Fandrich}\ \emph {et~al.}(2009)\citenamefont
  {Fandrich}, \citenamefont {Meinhardt},\ and\ \citenamefont
  {Grigorieff}}]{Fandrich2009}%
  \BibitemOpen
  \bibfield  {author} {\bibinfo {author} {\bibfnamefont {M.}~\bibnamefont
  {Fandrich}}, \bibinfo {author} {\bibfnamefont {J.}~\bibnamefont {Meinhardt}},
  \ and\ \bibinfo {author} {\bibfnamefont {N.}~\bibnamefont {Grigorieff}},\
  }\href@noop {} {\bibfield  {journal} {\bibinfo  {journal} {Prion},\ }\textbf
  {\bibinfo {volume} {3}},\ \bibinfo {pages} {89} (\bibinfo {year}
  {2009})}\BibitemShut {NoStop}%
\bibitem [{\citenamefont {Baxter}(2002)}]{Baxter-book}%
  \BibitemOpen
  \bibfield  {author} {\bibinfo {author} {\bibfnamefont {R.~J.}\ \bibnamefont
  {Baxter}},\ }\href@noop {} {\emph {\bibinfo {title} {Exactly solved models in
  statistical mechanics}}}\ (\bibinfo  {publisher} {Academic Press},\ \bibinfo
  {year} {2002})\BibitemShut {NoStop}%
\bibitem [{\citenamefont {Meakin}(1983)}]{Meakin1983}%
  \BibitemOpen
  \bibfield  {author} {\bibinfo {author} {\bibfnamefont {P.}~\bibnamefont
  {Meakin}},\ }\Doi {10.1103/PhysRevLett.51.1119} {\bibfield  {journal}
  {\bibinfo  {journal} {Phys. Rev. Lett.},\ }\textbf {\bibinfo {volume} {51}},\
  \bibinfo {pages} {1119} (\bibinfo {year} {1983})}\BibitemShut {NoStop}%
\bibitem [{gra()}]{grant-jack-prep}%
  \BibitemOpen
  \bibinfo {note} {J. Grant and R. L. Jack, in preparation}\BibitemShut
  {NoStop}%
\bibitem [{Note2()}]{Note2}%
  \BibitemOpen
  \bibinfo {note} {We also tested a modified scheme in which particles making
  multiple bonds in one move contribute ${\protect \mathcal {N}}_{\protect \rm
  new} - {\protect \mathcal {N}}_{\protect \rm old}$ to $K_+$, etc. The results
  from this scheme and the one used in the main text are essentially
  indistinguishable.}\BibitemShut {Stop}%
\bibitem [{\citenamefont {Sanz}\ \emph {et~al.}(2008)\citenamefont {Sanz},
  \citenamefont {Valeriani}, \citenamefont {Vissers}, \citenamefont {Fortini},
  \citenamefont {Leunissen}, \citenamefont {van Blaaderen}, \citenamefont
  {Frenkel},\ and\ \citenamefont {Dijkstra}}]{Sanz2008}%
  \BibitemOpen
  \bibfield  {author} {\bibinfo {author} {\bibfnamefont {E.}~\bibnamefont
  {Sanz}}, \bibinfo {author} {\bibfnamefont {C.}~\bibnamefont {Valeriani}},
  \bibinfo {author} {\bibfnamefont {T.}~\bibnamefont {Vissers}}, \bibinfo
  {author} {\bibfnamefont {A.}~\bibnamefont {Fortini}}, \bibinfo {author}
  {\bibfnamefont {M.~E.}\ \bibnamefont {Leunissen}}, \bibinfo {author}
  {\bibfnamefont {A.}~\bibnamefont {van Blaaderen}}, \bibinfo {author}
  {\bibfnamefont {D.}~\bibnamefont {Frenkel}}, \ and\ \bibinfo {author}
  {\bibfnamefont {M.}~\bibnamefont {Dijkstra}},\ }\href@noop {} {\bibfield
  {journal} {\bibinfo  {journal} {J. Phys.: Condens. Matt.},\ }\textbf
  {\bibinfo {volume} {20}},\ \bibinfo {pages} {494247} (\bibinfo {year}
  {2008})}\BibitemShut {NoStop}%
\bibitem [{\citenamefont {Garrahan}\ \emph {et~al.}(2009)\citenamefont
  {Garrahan}, \citenamefont {Jack}, \citenamefont {Lecomte}, \citenamefont
  {Pitard}, \citenamefont {van Duijvendijk},\ and\ \citenamefont {van
  Wijland}}]{Garrahan2009-kinks}%
  \BibitemOpen
  \bibfield  {author} {\bibinfo {author} {\bibfnamefont {J.~P.}\ \bibnamefont
  {Garrahan}}, \bibinfo {author} {\bibfnamefont {R.~L.}\ \bibnamefont {Jack}},
  \bibinfo {author} {\bibfnamefont {V.}~\bibnamefont {Lecomte}}, \bibinfo
  {author} {\bibfnamefont {E.}~\bibnamefont {Pitard}}, \bibinfo {author}
  {\bibfnamefont {K.}~\bibnamefont {van Duijvendijk}}, \ and\ \bibinfo {author}
  {\bibfnamefont {F.}~\bibnamefont {van Wijland}},\ }\Doi {ARTN 075007}
  {\bibfield  {journal} {\bibinfo  {journal} {J. Phys. A},\ }\textbf {\bibinfo
  {volume} {42}},\ \bibinfo {pages} {075007} (\bibinfo {year}
  {2009})}\BibitemShut {NoStop}%
\end{thebibliography}
\end{document}